\documentclass[12pt]{article}
\usepackage{geometry}
\usepackage[switch]{lineno}
\usepackage{setspace} 

\usepackage{amsmath, amssymb, graphicx,fullpage,color,mathtools,amsthm,xcolor}
\usepackage{caption, subcaption}
\usepackage{array}
\usepackage{url}
\usepackage{tikz}
\usepackage{tabularx}
\usepackage{booktabs}
\usepackage{cite}

\usepackage{microtype}
\usepackage{hyperref,color}

\definecolor{webgreen}{rgb}{0,.35,0}
\definecolor{webbrown}{rgb}{.6,0,0}
\definecolor{RoyalBlue}{rgb}{0,0,0.9}
\definecolor{purp}{rgb}{0.6,0.05,0.8}
\definecolor{ora}{rgb}{0.7,0.35,0.02}

\begin{document}

\doublespacing

\title{Robust Parametric Estimation of Avian Cranial Morphology}
\author{Kaikwan Lau$^{1}$, Gary P. T. Choi$^{1,\ast}$\\
\footnotesize{$^{1}$Department of Mathematics, The Chinese University of Hong Kong}\\
\footnotesize{$^\ast$To whom correspondence should be addressed; E-mail: ptchoi@cuhk.edu.hk}
}
\date{ }

\maketitle

\begin{abstract}
Understanding the growth and form of complex morphological structures is one of the most fundamental problems in biology. While many prior works have analyzed the beak morphology of Darwin's finches, other cranial features are relatively less explored. In this work, we develop geometric and statistical methods for analyzing the skull morphology of Darwin's finches and their relatives, focusing on the relationship between their skull dimensions, orbit curvature, and neurocranial geometries. Unlike traditional landmark-based approaches that scale linearly with human labor, our framework is fully unsupervised. Specifically, by utilizing tools in computational geometry, differential geometry, and numerical optimization, we develop efficient algorithms for quantifying various key geometric features of the skull. We then perform a statistical analysis and discover a strong correlation between skull size and orbit curvature. Based on our findings, we further establish a predictive model that can estimate the orbit curvature using easily obtainable linear skull measurements. Our results show that the predictive model is highly effective and capable of explaining 85.48\% of the variance in curvature with an average prediction error of only 6.35\%. Altogether, our work establishes a rigorous foundation for the digital estimation and high-throughput phenotyping of large-scale museum collections, overcoming the scalability bottlenecks of manual methods.
\end{abstract}

\section{Introduction}

The morphology of the avian skull offers critical insights into the evolutionary adaptations of birds to their respective ecological niches~\cite{bock1974avian}. In particular, the avian skull is a highly integrated functional complex of bones rather than a monolithic structure. The modular variation within these specific elements, such as the beak, neurocranium, and orbits, and how they respond to interconnected evolutionary pressures~\cite{zusi1993patterns}, is what ultimately drives the remarkable morphological diversity observed across avian lineages. Avian skull evolution is therefore not simply a reflection of feeding ecology, but a complex integration shaped by biomechanical constraints and factors such as habitat density and migration~\cite{hunt2023ecological}. Darwin's finches, a classic model for adaptive radiation, have been extensively studied~\cite{grant201440,farrington2014evolutionary,navalon2020consequences}. In particular, most prior works have focused on the morphology of the beaks of Darwin's finches and their implications~\cite{herrel2005bite,grant2006evolution,al2021geometry,mosleh2023beak}. Besides the finch beaks, Genbrugge et~al.~\cite{genbrugge2011head} analyzed the head length, depth, and width of \emph{Geospiza fortis}. Also, Tokita et~al.~\cite{tokita2017cranial} compared the cranial morphology of Darwin's finches and Hawaiian honeycreepers using landmark-based comparative morphometrics. However, many other cranial features of Darwin's finches remain less studied. For other avian skulls, a few prior works have explored the relationship between the avian eye, orbit, and sclerotic ring~\cite{hall2008anatomical}, the relationship between brain shape and orbital shape~\cite{kawabe2013variation}, and the exo- and endocranial covariation~\cite{marugan2006avian}. Therefore, it is natural to ask whether one can extract more geometric features from the skulls of Darwin's finches and their relatives and perform a more comprehensive analysis on the relationship between different features. 
While high-dimensional shape analysis techniques such as landmark-based geometric morphometrics and diffeomorphic mappings are commonly used, they often impose strict topological requirements on the given shape data. Natural fenestrae and taphonomic damage (holes, cracks) frequently render museum specimens topologically complex. Consequently, standard geometric morphometrics, diffeomorphic mapping, or parameterization methods often fail or require labor-intensive manual repair, thereby hindering the high-throughput analysis of large-scale datasets. Therefore, it is desirable to have a complementary, fully unsupervised framework that is capable of extracting robust parametric estimates from such imperfect data.

\begin{figure*}[t!]
    \centering
    \includegraphics[width=\linewidth]{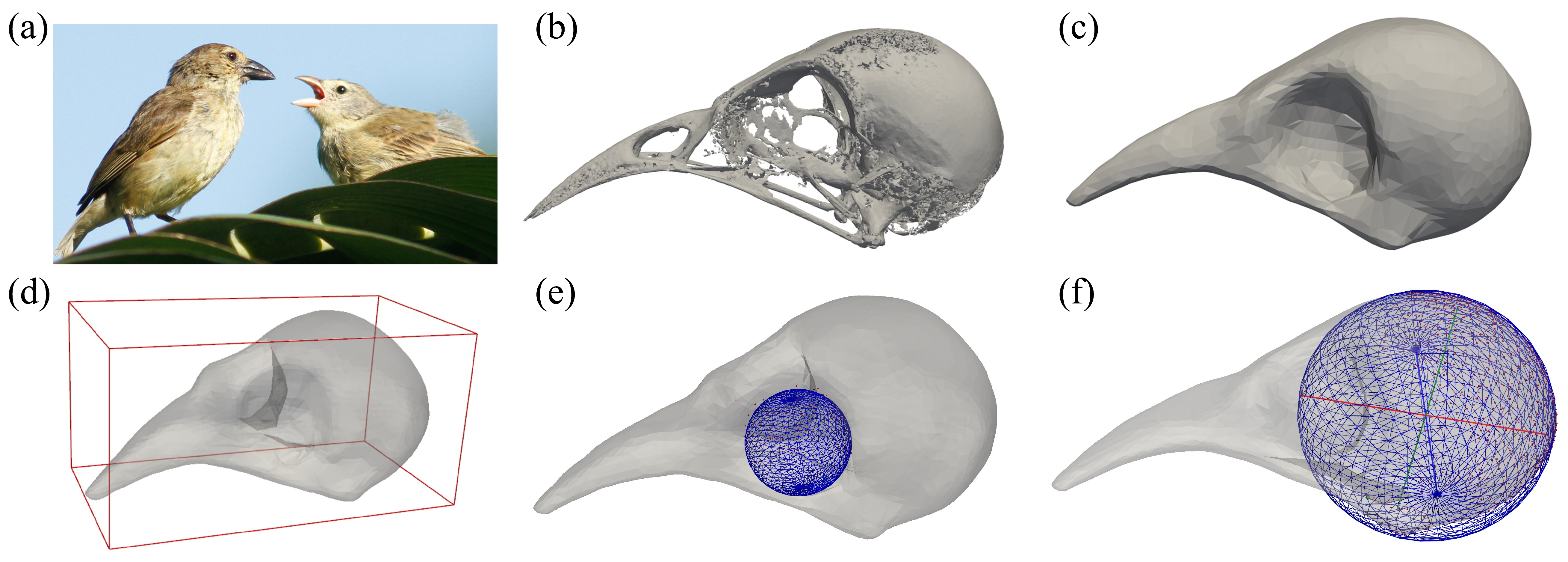}
    \caption{\textbf{An overview of the proposed skull shape quantification pipeline.} (a)~An example of Darwin’s finches (\emph{Camarhynchus pallidus}) obtained from Wikimedia Commons~\cite{wikimedia} under the CC BY 4.0 License. (b)~A raw skull mesh of a \emph{Pinaroloxias inornata} specimen in the dataset we considered. (c)~The preprocessed skull mesh with enhanced smoothness. (d)--(f)~Based on the preprocessed skull mesh, we can perform (d)~bounding box calculation for quantifying the skull dimensions, (e)~sphere fitting for quantifying the curvature of the orbits, and (f)~ellipsoid fitting for quantifying the geometry of the neurocranium (braincase).}
    \label{fig:overview}
\end{figure*}

Motivated by the above works, in this paper we develop geometric and statistical methods to analyze the skull morphology of Darwin's finches in terms of the relationship between the skull dimensions, the orbit curvature, and neurocranial geometries (see Fig.~\ref{fig:overview} for an overview). Specifically, using tools in computational geometry, differential geometry, and numerical optimization, we quantify various key geometric features from the 3D skull scans. To address the topological issues in the specimens, we adopt a strategy of methodological parsimony, utilizing robust geometric primitives that remain stable even in the presence of topological noise. We then perform statistical analyses to study the relationship between these features, from which we discover a strong correlation between the orbital shape and the skull dimensions. We further develop a predictive model for orbit curvature from more easily obtainable linear skull measurements, which not only reveal intriguing relationships between these geometric quantities but also potentially aid in the reconstruction and estimation of complex geometric features of other damaged or incomplete specimens. Altogether, our work provides a deeper, quantitative understanding of the functional and evolutionary pressures that have shaped the remarkable diversity of avian skull morphology.

\section{Materials and Methods}
To extract various geometric features from the 3D avian skulls, we develop several methods using tools in computational geometry, differential geometry, and numerical optimization.

\subsection{Dataset and Mesh Processing}
In this work, we analyze a dataset comprising 100 full skull specimens of Darwin's finches and their relatives. The raw micro-CT scans and preliminary mesh data for these specimens were originally acquired, processed, and described in~\cite{tokita2017cranial,al2021geometry,mosleh2023beak}. The original scans were taken using an XRA-002 micro-CT scanner at micron scale under the condition of 70 kV and 70 $\mu$A (see~\cite{tokita2017cranial} for details), and all skull dimension measurements reported in this work are expressed in millimeters (mm). The 100 specimens cover Darwin's finches in the genera \emph{Camarhynchus}, \emph{Certhidea}, \emph{Geospiza}, \emph{Pinaroloxias}, and \emph{Platyspiza}, and their relatives in the genera \emph{Coereba}, \emph{Euneornis}, \emph{Loxigilla}, \emph{Melopyrrha}, and \emph{Tiaris} (see SI Section~S1 for details). 

We first apply the voxelization and smoothing procedure as described in~\cite{mosleh2023beak} to remesh the original 3D scans (see Fig.~\ref{fig:overview}(b) for an example) with various hole-filling and cleaning steps. As shown in Fig.~\ref{fig:overview}(c), the processed mesh contains a large, contiguous component of vertices and faces with enhanced smoothness, and extraneous and disconnected artifacts are removed. This helps improve the robustness of the subsequent geometric curvature quantification and morphological fitting analysis. We remark that while the remeshing step includes preliminary hole-filling to address minor mesh artifacts, the remeshed scans still exhibit topological diversity. Therefore, our subsequent algorithm is designed to be robust to residual topological artifacts and surface noise. 

Also, we enforce that all specimens are aligned within a uniform coordinate system for easier comparison and analysis. To achieve this, all raw skull meshes in the dataset are first oriented based on their principal moment directions as described in~\cite{al2021geometry}. {Specifically, following the methodology detailed in the Supplementary Information of~\cite{al2021geometry}, the principal moment directions are defined as the eigen-directions of the moment tensor
\begin{equation}
    M_{ij}=\sum_{n=1}^{N}(P_{n}\cdot P_{n}\delta_{ij}-P_{ni}P_{nj}),
\end{equation}
which is calculated from all vertices $\{P_{n}\}_{n=1}^{N}$ of the 3D skull mesh, where $i,j\in\{1,2,3\}$ correspond to the $x, y, z$ coordinates, and $\delta_{ij}$ is the Kronecker delta. By computing these eigen-directions, the skulls are objectively aligned such that the direction of the highest moment is oriented along the $x$-axis, and the direction of the lowest moment is along the $z$-axis. Because eigen-decomposition inherently possesses sign ambiguity and potential axis-swapping, the moment tensor alignment alone may result in some meshes facing opposite directions or being misaligned by 90-degree increments. To resolve this, we further applied 90-degree or 180-degree rotations to the affected meshes. This step ensured a consistently oriented anatomical coordinate system across the entire dataset prior to bounding box calculation, where the $x$, $y$, and $z$-axes universally correspond to the anatomical length, width, and height. This standardized orientation facilitates our quantification of the skull dimensions, which will be described in detail in the following section.

\subsection{Linear Measurements of Skull Dimensions}

From the processed meshes, we extract three key linear measurements of the skull, including the length $x$, width $y$, and height $z$, using a bounding box calculation approach (see Fig.~\ref{fig:overview}(d)). To ensure our linear measurements were consistent and biologically meaningful, we employ the Axis-Aligned Bounding Box (AABB) approach~\cite{bergen1997efficient}, which requires all specimens to be pre-aligned to a uniform coordinate system (achieved via the moment tensor orientation and orthogonal corrections described in the previous section), where the $x$, $y$, and $z$-axes correspond to the anatomical length, width, and height, respectively. Note that one may also consider using the Oriented Bounding Box (OBB) approach~\cite{gottschalk1996obbtree}, which allows the bounding box to rotate to better fit the object. However, we find that the OBB approach is less suitable for the geometric quantification of the skulls in our problem, as the skull may be tilted in the resulting OBB in some cases, leading to inconsistencies in the definition of length, width, and height in our analysis. For instance, as shown in Fig.~\ref{fig:parallel_bounding_boxes}(a), the ``length'' obtained from the OBB approach does not represent the dimension of the skull correctly. By contrast, the AABB approach (Fig.~\ref{fig:parallel_bounding_boxes}(b)) produces an axis-aligned bounding box, thereby giving more interpretable and meaningful measurements of the skull dimensions. Hence, the AABB approach is used in our study. 

\begin{figure}[t]
    \centering
    \includegraphics[width=\linewidth]{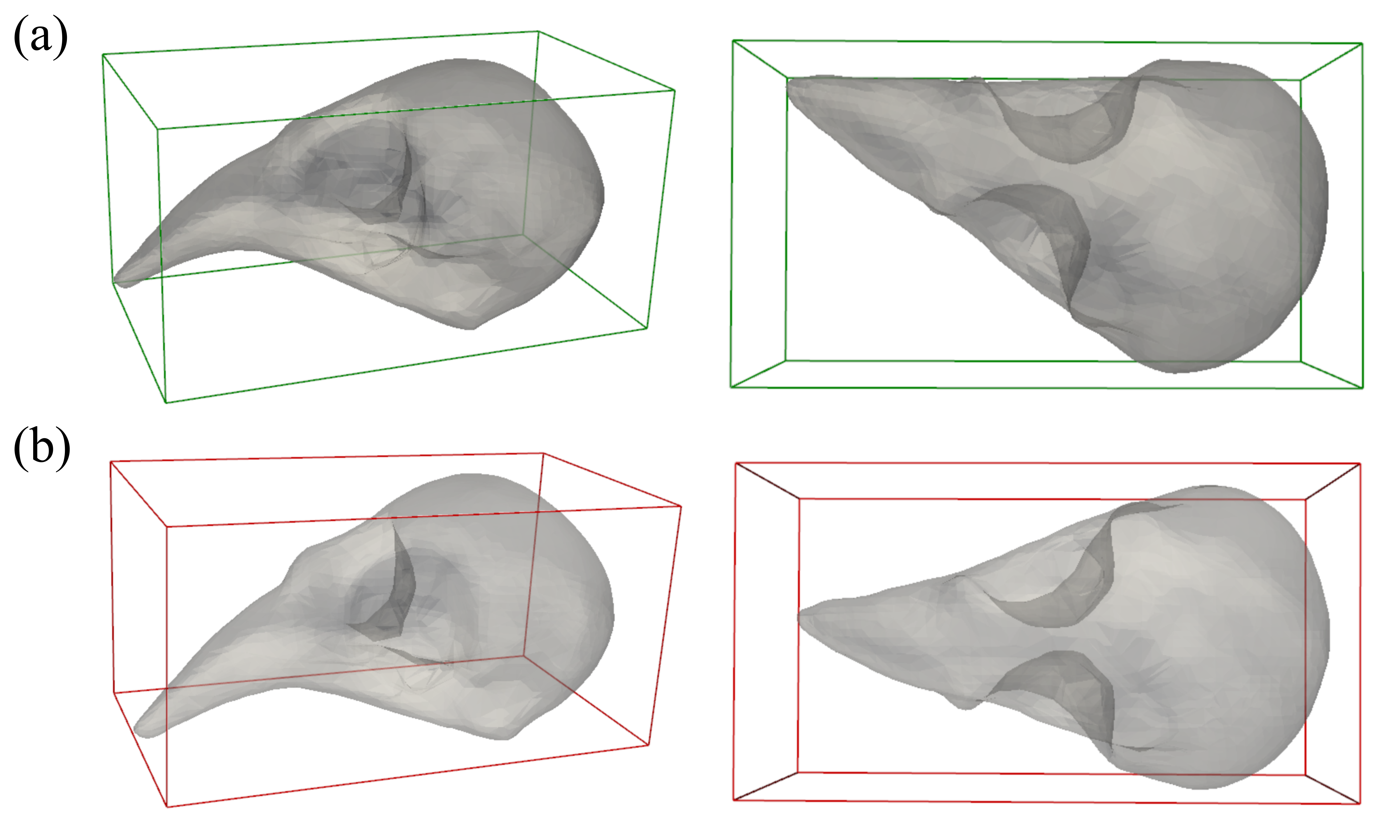}
    \caption{\textbf{Comparison between bounding box calculation methods for quantifying the skull dimensions of a \emph{Geospiza septentrionalis} skull.} (a) The Oriented Bounding Box (OBB) result with two different views. (b) The Axis-Aligned Bounding Box (AABB) result with two different views.}
    \label{fig:parallel_bounding_boxes}
\end{figure}

\begin{figure*}[t]
    \centering
    \includegraphics[width=\linewidth]{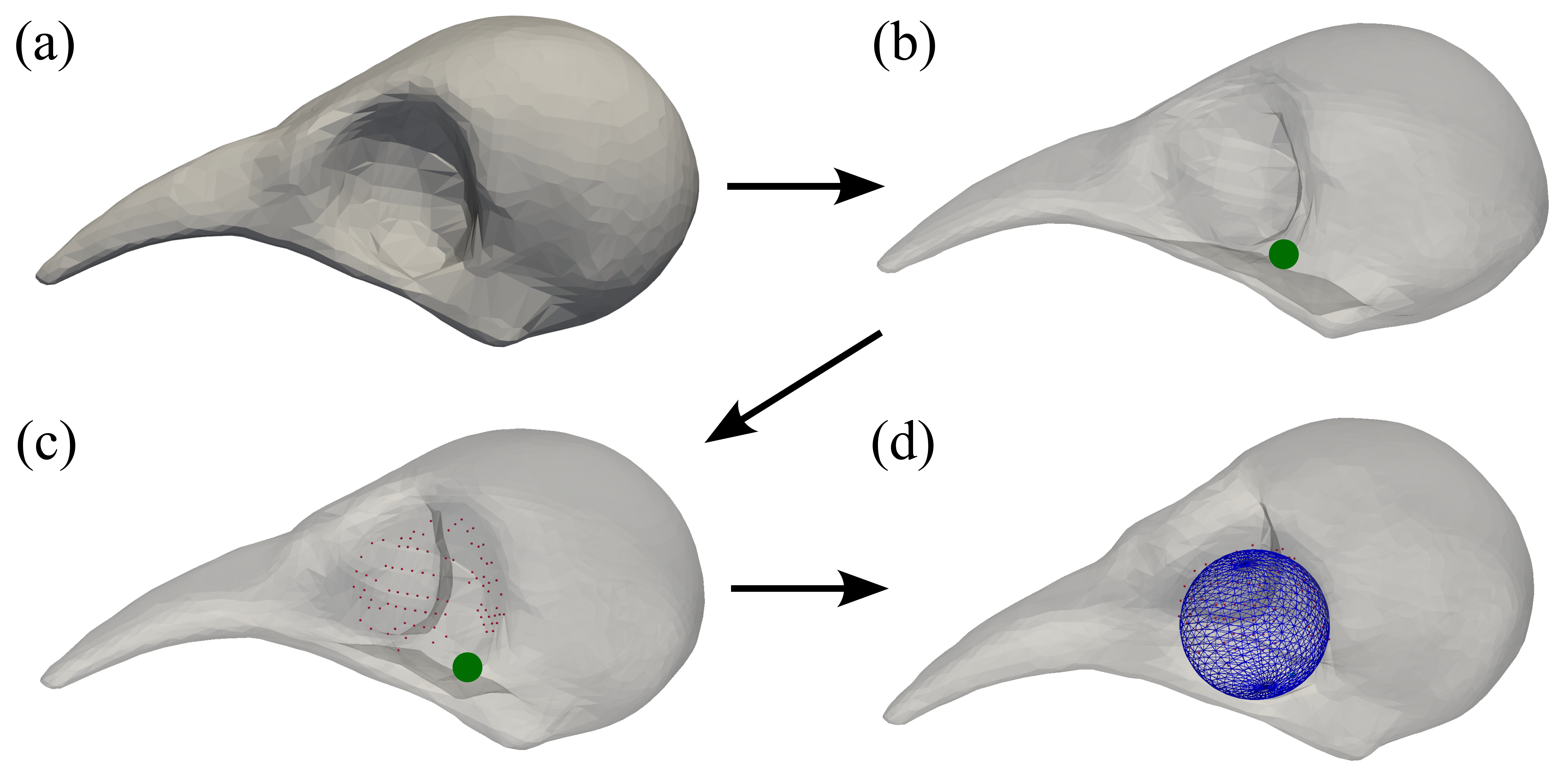}
    \caption{\textbf{An illustration of the robust and iterative sphere fitting method.} (a)~A 3D skull mesh of \emph{Pinaroloxias inornata}. (b) A seed point (green) is found to locate the orbital socket. (c) Selected points (red) are then identified via connected component identification and distance-based thresholding. (d) An optimal sphere (blue) is finally obtained via an optimization problem for approximating the curvature of the orbital socket.}
    \label{fig:orbit_fitting}
\end{figure*}

\subsection{Automatic identification of the orbit region}

Next, we utilize differential geometry for the automatic identification of the orbit region for our subsequent analysis. Specifically, note that the orbital socket is a region in the skull characterized by significant concavity. Motivated by this observation, for every vertex in the skull mesh, we compute the mean curvature 
\begin{equation}
    H = \frac{1}{2}(\kappa_1+\kappa_2),
\end{equation}
where $\kappa_1, \kappa_2$ are the principal curvatures. To focus the analysis on the orbit, we restrict the search to a region of interest (ROI) corresponding to the middle 40\% of the skull along its longitudinal (length) axis. This relative band is applied uniformly across all specimens, scaling to each specimen's own length. Within this band, we identify the seed point as the single most concave vertex, i.e., the vertex of globally minimum mean curvature, which requires no hand-tuned curvature threshold. Starting from this seed, we grow a connected component of vertices with pronounced negative mean curvature to delineate a single, contiguous concave patch corresponding to one orbital socket. Here, note that the mean curvature was selected as our primary metric as it measures how much a surface bends at a specific point. This makes it exceptionally effective at identifying concave regions like the orbital socket, which can be thought of as a ``dent'' in the skull's surface, allowing for reliable identification of our region of interest. From this pool of candidates, a singular ``seed point'' is established, corresponding to the vertex that possesses the absolute minimum curvature value. See Fig.~\ref{fig:orbit_fitting}(a)--(b) for an illustration of the above processes.

We then utilize a connected component identification approach to ensure that the analytical focus remains on a singular, coherent anatomical feature (see Fig.~\ref{fig:orbit_fitting}(c)). Specifically, a contiguous cluster of points that corresponds to a single orbital socket is isolated. To achieve this, we first construct a graph from the candidate points. Then, the connected component containing the above-mentioned seed point is extracted. However, it is noteworthy that the mean curvature computation relies on the discretization of the mesh and may be affected by minor variations of the vertices locally. To further obtain a unified quantity representing the orbit curvature, in the subsequent section we develop a sphere fitting algorithm based on the points extracted from the connected component.

\subsection{Robust and Iterative Sphere Fitting}
To quantify the functional envelope of the orbit while rejecting surface noise, we develop a robust and iterative sphere fitting algorithm (Fig.~\ref{fig:overview}(e)). Specifically, the fitting procedure is formulated as an optimization problem that seeks to minimize a loss function, defined as the sum of squared Euclidean distances from each point to the surface of a candidate sphere. To enhance the robustness of the fit, an iterative scheme is developed. 

As an outline of the iterative scheme, we start with the automatic identification of the orbit region as described in the previous section. Then, we solve an optimization problem to fit a sphere to the current set of candidate points of the orbit region. We then use the distance of each candidate point from the fitted sphere to identify outlier points. By removing the outlier points, we obtain a refined set of candidate points representing the orbit region, from which we can repeat the above-mentioned sphere fitting procedure. This iterative outlier-removal process ensures that the final sphere parameters yield a more precise and representative geometric model of the feature. The detailed mathematical formulation and computational procedure for this iterative process are explained as follows.

Let $P = \{p_1, p_2, \dots, p_n\}$, where each point $p_i = (x_i, y_i, z_i)$ is a vector in $\mathbb{R}^3$. A sphere is defined by its center $C = (c_x, c_y, c_z)$ and its radius $r$. The objective is to find the parameters $(C, r)$ that best fit the points in $P$ (see Fig.~\ref{fig:orbit_fitting}(d)). The distance of a point $p_i$ from the center $C$ is given by the Euclidean norm:
\begin{equation}
d_i = \|p_i - C\| = \sqrt{(x_i - c_x)^2 + (y_i - c_y)^2 + (z_i - c_z)^2}.
\end{equation}

The error for a single point, $e_i$, is then defined as the absolute difference between its distance from the center and the sphere's radius:
\begin{equation}
e_i = |d_i - r| = |\|p_i - C\| - r|.
\end{equation}

In other words, $e_i$ represents the shortest distance from the point to the surface of the sphere.

Now, since the objective is to find the sphere parameters that best fit the points in $P$, we define the loss function $\mathcal{L}(C,r)$ as the sum of the squared errors for all points in the point set $P$:
\begin{equation}
\mathcal{L}(C, r) = \sum_{i=1}^{n} e_i^2 = \sum_{i=1}^{n} \left( \|p_i - C\| - r \right)^2.
\end{equation}

Then, the optimization problem is to find the parameters that minimize this function:
\begin{equation}
    (\hat{C}, \hat{r}) = \underset{C \in \mathbb{R}^3, r \in \mathbb{R}^+}{\text{argmin}} \mathcal{L}(C, r).
\end{equation}

To solve the above-mentioned optimization problem of fitting a sphere to the data points, we use the Limited-memory Broyden--Fletcher--Goldfarb--Shann (L-BFGS-B) algorithm~\cite{byrd1995limited} {\color{blue} from the \texttt{scipy} library in Python} as it is a quasi-Newton method highly adept at solving non-linear optimization problems. This approach is computationally efficient and, critically, allows for the enforcement of bound constraints, which is essential for ensuring a physically realistic solution by requiring the sphere's radius to be a positive value. The algorithm terminates when the projected gradient of the loss function reaches a prescribed convergence tolerance (set as $10^{-5}$ in our experiments). This is followed by the identification and removal of outlier points, which are defined as points whose distance from the sphere's surface exceeds a predetermined threshold of 2.0 standard deviations above the mean error. We then repeat the above sphere fitting optimization procedure with the outlier points removed. This robust refinement loop continues until the fitting result stabilizes, with no further vertices being identified as outliers. For better control of the computation, one may also terminate the iteration if a prescribed maximum number of refinement steps is reached or the point cloud size drops below 4 points, the minimum required to uniquely determine the sphere parameters $(c_x, c_y, c_z, r)$. In practice, we find that a maximum of three refinement steps is sufficient to yield stable results for our subsequent analysis. Finally, the orbit curvature is computed by $1/\hat{r}$, where $\hat{r}$ is the sphere radius. In Fig.~\ref{fig:sphere_examples}, we show several examples of the sphere fitting results, from which we can see that the method is capable of quantifying the orbit curvature for different skull specimens. More examples are provided in SI Section~S1.

To facilitate the biological interpretation of these numerical parameters, Fig.~\ref{fig:difference_curvature} provides a visual comparison of morphological extremes within the dataset. While our primitive-fitting approach does not produce continuous deformation grids characteristic of traditional geometric morphometrics, mapping the robust parameters back onto the 3D meshes clearly illustrates the physical manifestations of the statistical data. For instance, contrasting the high orbit curvature of \emph{Tiaris canora} (Fig.~\ref{fig:difference_curvature}(a)) against the low orbit curvature of \emph{Geospiza conirostris} (Fig.~\ref{fig:difference_curvature}(b)) explicitly reveals how the variation in the calculated sphere radius corresponds to tangible differences in the spatial packing of the cranial architecture. This graphical representation ensures that the parametric output remains grounded in observable comparative anatomy.

\begin{figure}[t]
    \centering
    \includegraphics[width=\linewidth]{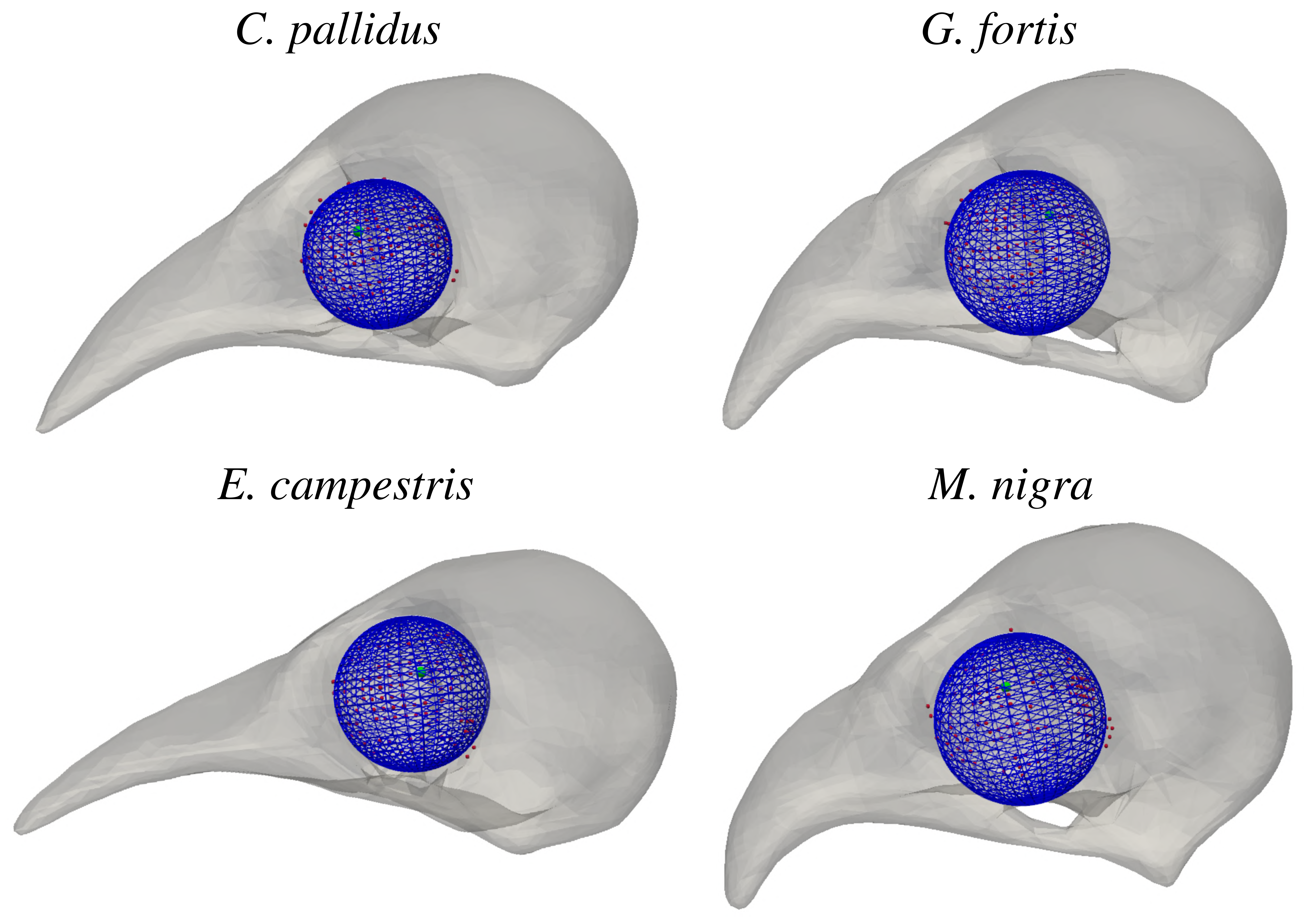}
    \caption{\textbf{Four examples of the sphere fitting results for orbit curvature quantification}. (a) A skull of \emph{Tiaris canora} (specimen T.canoraE) exhibiting high orbit curvature (smaller fitted sphere). (b) A skull of \emph{Geospiza conirostris} (specimen G.ConirostrisD) exhibiting low orbit curvature (larger fitted sphere), demonstrating a broader, more open orbital architecture. The red dots represent the inlier surface points used for the robust algorithmic fitting.}
    \label{fig:sphere_examples}
\end{figure}

\begin{figure}[t]
    \centering
    \includegraphics[width=\linewidth]{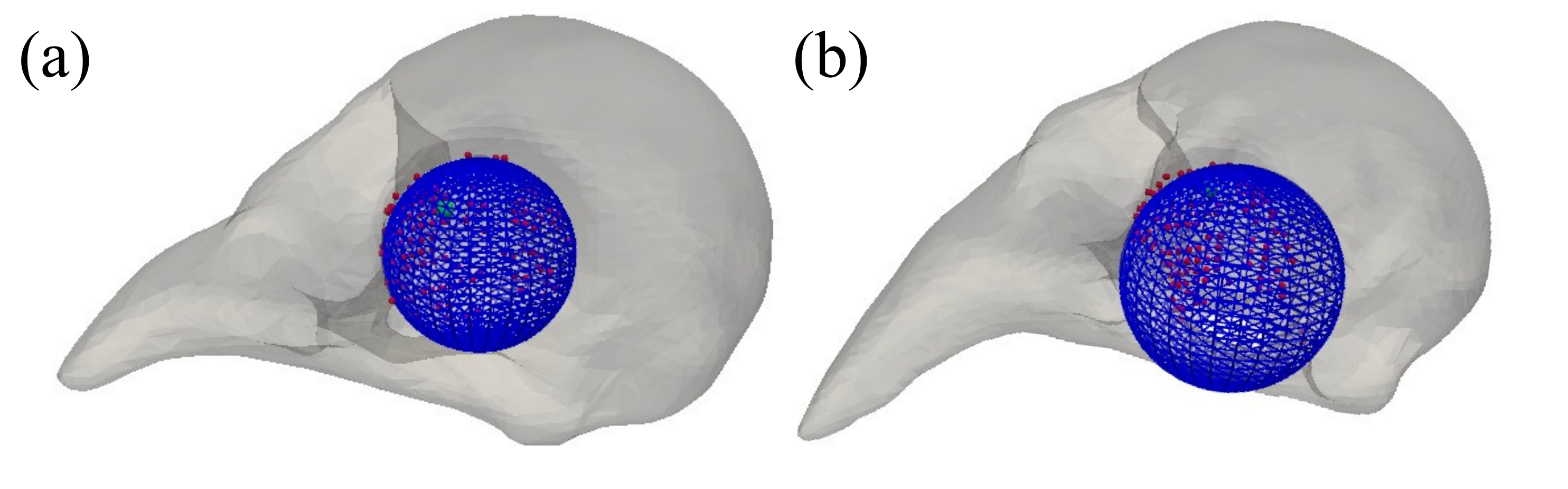}
    \caption{\textbf{Visualizing morphological variation through parametric geometric primitives.} (a) A specimen exhibiting high orbit curvature (smaller fitted sphere), illustrating a tighter, more concave orbital socket constraint. (b) A specimen exhibiting low orbit curvature (larger fitted sphere), demonstrating a broader, more open orbital architecture. The red dots represent the inlier surface points used for the robust algorithmic fitting.}
    \label{fig:difference_curvature}
\end{figure}

\subsection{Extension to Ellipsoid Fitting of Neurocranium}
By extending the above-mentioned sphere fitting algorithm, we can achieve an ellipsoid fitting of the neurocranium of the skull and quantify the shape of the braincase using the ellipsoidal semi-axis lengths $a, b, c$ (see Fig.~\ref{fig:ellipsoid_fitting} for an illustration).

\begin{figure}[t!]
    \centering
    \includegraphics[width=\linewidth]{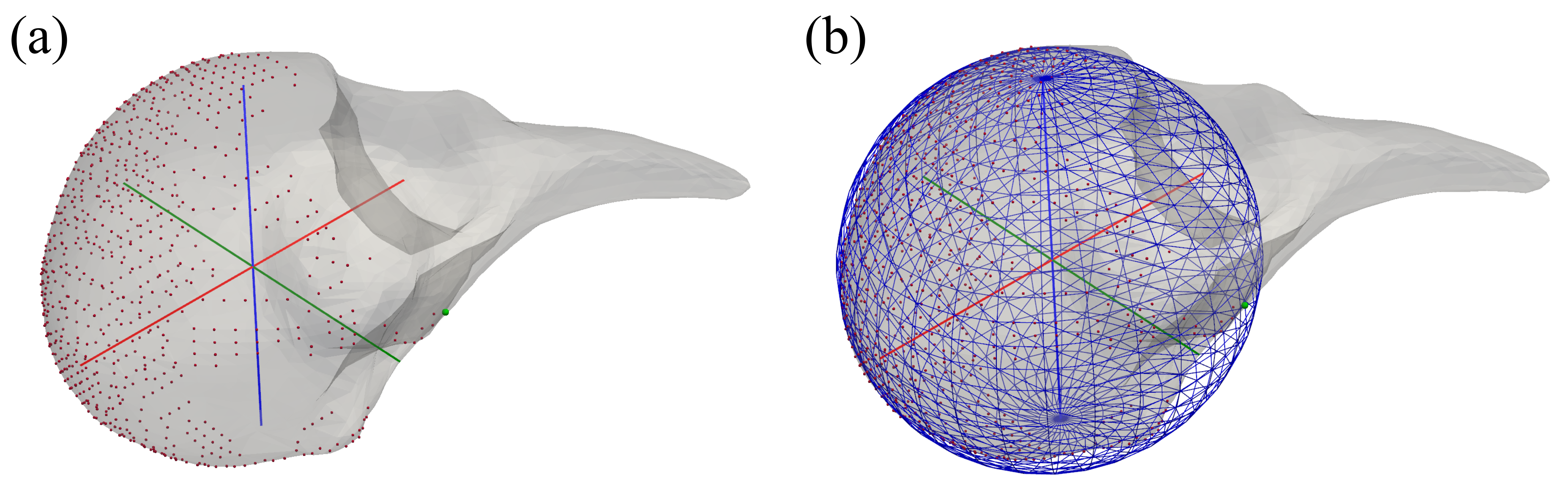}
    \caption{\textbf{An illustration of the ellipsoid fitting algorithm for a \emph{Pinaroloxias inornata} skull}. (a) The input skull mesh with the seed point (green dot), selected points (red dots), and the computed axis $a$ (red line), $b$ (green line), and $c$ (blue line). (b) The best-fit ellipsoid obtained by the proposed algorithm.}
    \label{fig:ellipsoid_fitting}
\end{figure}

Specifically, to ensure that the ellipsoidal fitting analysis is performed on the neurocranium of the skull, we first calculate a cutoff coordinate for isolating the posterior region of the 3D skull mesh based on a prescribed threshold. We then select all vertices in the mesh that are beyond this cutoff, effectively isolating the rearmost 40\% of the skull. This area becomes the ROI for our neurocranial shape quantification. Within this ROI, we search for the single most outwardly curved (convex) point and designate it as the ``seed point''. This provides a reliable starting anchor on the roundest part of the neurocranium. We further refine the selection to get the best possible set of points. Specifically, a large group of the most convex points within the ROI is then identified based on a prescribed parameter. To avoid using scattered points, we adopt a similar strategy as in the previously described orbit analysis and perform a connected component identification procedure. Starting from the seed point, we find all the points from the ``curved'' cloud that are connected to it. This results in a single, coherent patch of points on the surface of the neurocranium, which is an ideal input for the ellipsoid fitting algorithm. It is worth noting that the core curvature-seeking algorithm is intrinsically robust and successfully isolates the neurocranium for the vast majority of the dataset without requiring this spatial constraint. However, this conservative 40\% relative length threshold was implemented as an automated safeguard to handle the significant topological artifacts prevalent in museum scans. Several specimens in the dataset exhibit severe taphonomic damage, fragmentation, or anomalous internal fenestrations within the cranial cavity. By restricting the ROI, we ensure the algorithm avoids these highly degraded regions, allowing the pipeline to reliably process all meshes in the dataset without requiring manual repair or specimen exclusion.

Similar to the sphere fitting algorithm for the orbit, we take the clean patch of points and find the best possible axis-aligned ellipsoid to fit them. More specifically, here we search for the optimal ellipsoid centers $(c_x, c_y, c_z)$ and semi-axis lengths $a,b,c$ that minimize the following function:
\begin{equation}
\mathcal{L}_{\text{ellipsoid}}(c_x,c_y,c_z, a,b,c) = \sum_{i=1}^{n} \left( \frac{(x_i-c_x)^2}{a^2}+ \frac{(y_i-c_y)^2}{b^2} + \frac{(z_i-c_z)^2}{c^2} - 1 \right)^2,
\end{equation}
where $\{(x_i, y_i, z_i)\}_{i=1}^n$ are the identified points. Note that each term in $\mathcal{L}_{\text{ellipsoid}}$ measures how far a point is from the surface of the guessed ellipsoid, and hence minimizing $\mathcal{L}_{\text{ellipsoid}}$ gives the best-fit ellipsoid to the set of identified points. To solve this optimization problem, we again use the L-BFGS-B method from the \texttt{scipy} library in Python with the prescribed convergence tolerance of $10^{-5}$. Moreover, analogous to the sphere fitting algorithm, we further identify any points that lie too far from the fitted ellipsoid surface (``outliers''), which are defined as
points whose distance from the ellipsoid surface exceeds 1.5 standard deviations above the mean error. We then remove these outliers and run the entire optimization process again on the cleaner set of points, thereby making the result less sensitive to noise. As with the sphere fitting algorithm, the iterative process terminates when the result stabilizes, with no further vertices being identified as outliers. We also terminate the process if a maximum of three refinement steps is reached or the point cloud size drops below 6 points, the minimum required to uniquely determine the ellipsoid parameters $(c_x, c_y, c_z, a,b,c)$. In Fig.~\ref{fig:ellipsoid_examples}, we show several examples of the ellipsoid fitting results, from which it can be observed that the algorithm handles different neurocranial geometries very well. More examples are provided in SI Section~S1.

\begin{figure}[t]
    \centering
    \includegraphics[width=\linewidth]{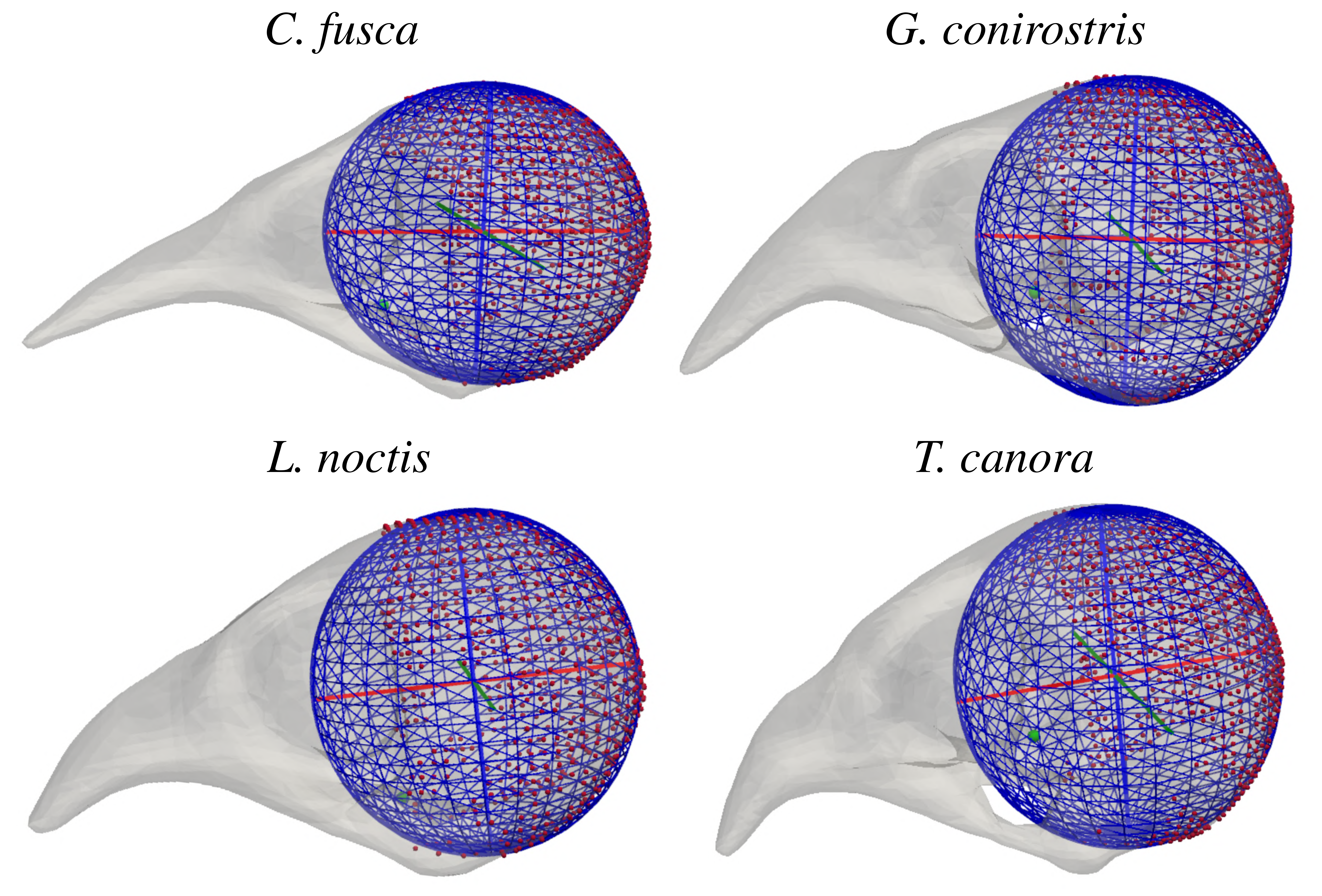}
    \caption{\textbf{Four examples of the ellipsoid fitting results for neurocranial geometry quantification.} In each example, the seed point is highlighted in green, and the identified points for the fitting are highlighted in red. The optimal ellipsoid is displayed in blue. Note that the figures are not displayed to scale.}
    \label{fig:ellipsoid_examples}
\end{figure}

\section{Results}

\subsection{Correlation between Skull Dimensions and Orbit Curvature}

After applying the proposed approaches for extracting the skull dimension and geometric curvature parameters, we perform a detailed analysis of the quantities. In Fig.~\ref{fig:plots_with_data_and_tree}, we present the phylomorphospace of the orbit curvature in Darwin's finches and their relatives, which illustrates the wide range of morphological variation across both the finch radiation and the outgroup relatives.

\begin{figure*}[t]
    \centering
    \includegraphics[width=\linewidth]{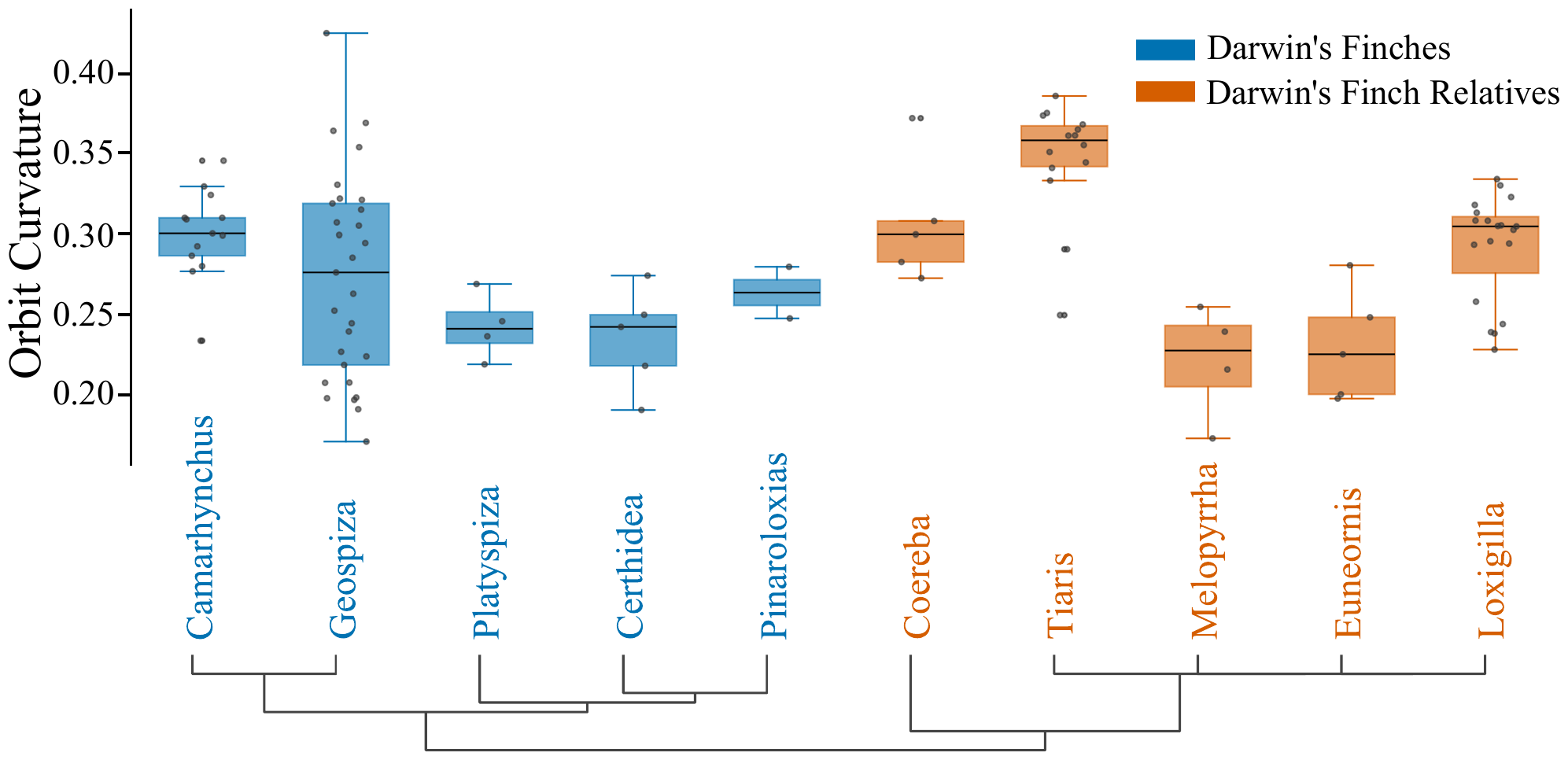}
    \caption{\textbf{Phylomorphospace of orbit curvature in Darwin’s finches and their relatives.} (Top) Box plots illustrating the variation in orbit curvature across ten genera. The central line represents the median, the box limits indicate the interquartile range (IQR), and the whiskers extend to 1.5 $\times$ IQR. Individual data points are overlaid to display the distribution of the 100 specimens analyzed. (Bottom) A consensus phylogenetic tree showing the evolutionary relationships among the genera. The branching topology is based on genomic consensus~\cite{lamichhaney2015evolution, burns2014phylogenetics}, placing \emph{Certhidea} as the basal lineage of Darwin's finches and identifying \emph{Geospiza} and \emph{Camarhynchus} as recently diverged sister groups. The alignment highlights the phylogenetic distribution of orbital shape, illustrating the wide range of morphological variation across both the finch radiation (blue boxes) and the outgroup relatives (orange boxes).}
    \label{fig:plots_with_data_and_tree}
\end{figure*}

\begin{figure*}[t!]
    \centering
    \includegraphics[width=1\linewidth]{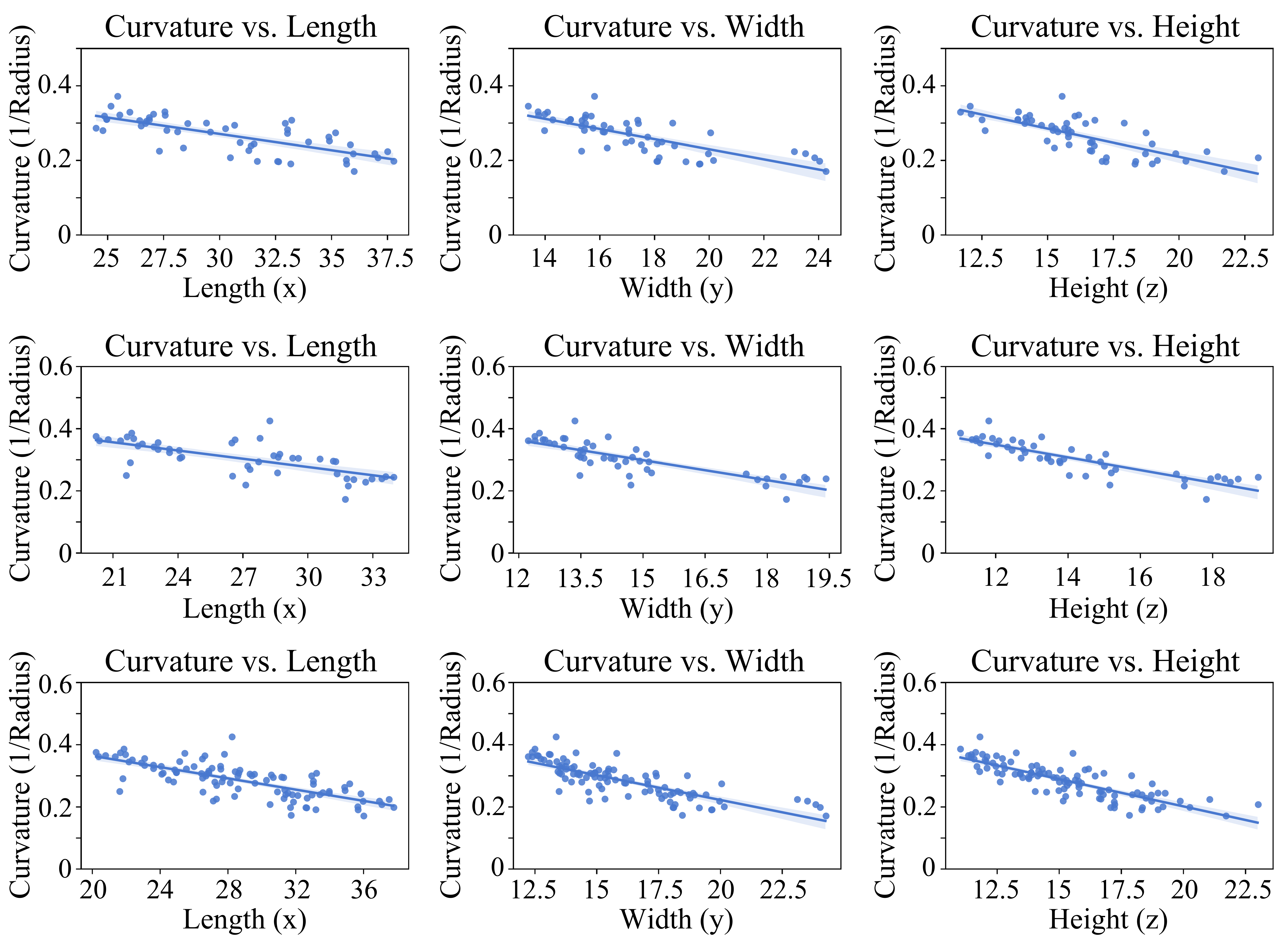}
    \caption{\textbf{Correlation between different geometric parameters of the skulls of Darwin’s finches and their relatives.} Each plot shows the correlation between the orbit curvature against an object dimension parameter (length/width/height) for (a) Darwin’s finches, (b) their relatives, and (c) both groups combined. The blue line represents the regression line in each plot, and the shaded region around each regression line represents the confidence interval for the linear approximation.}
    \label{fig:plots_with_regression}
\end{figure*}
Next, we perform a statistical analysis on the orbit curvature and the skull dimensions. Our analysis reveals a strong correlation between the dimensions of the skull and the curvature of the orbit (see Fig.~\ref{fig:plots_with_regression}). For both Darwin's finches and their relatives, a high Spearman correlation is observed between the sphere radius (inversely related to the orbit curvature) and the skull's length, width, and height. Specifically, for Darwin's finches, the Spearman correlation coefficients between sphere radius and length $x$, width $y$, and height $z$ are 0.740, 0.802, and 0.774, respectively. For the relatives of Darwin's finches, these correlations are even stronger: 0.708, 0.843, and 0.893. This indicates that larger skulls tend to have larger orbit radii, and therefore, lower curvature. The Pearson correlation analysis confirms this inverse relationship, showing negative correlations between curvature and the skull dimensions. If we consider all Darwin's finches and their relatives together, the correlation coefficients are 0.758, 0.832, and 0.844. It is noteworthy that these findings align with established principles of biological scaling, or allometry, where the relative size of anatomical features changes with overall body size~\cite{schmidt1984scaling}.

Also, note that the shaded region around each regression line in Fig.~\ref{fig:plots_with_regression} represents the confidence interval for the linear approximation. For the main cluster of data points (curvature between 0.2 and 0.4), this interval is relatively narrow, indicating that the linear model is a reasonable approximation for describing the central tendency of the data. More detailed statistical results are provided in SI Section~S2.

\subsection{Curvature Modeling and Prediction}

As shown in the analysis above, there is a strong linear relationship between the curvature of the orbit and the dimensions of the skull. To investigate the relationship between skull dimensions and orbit curvature further, we develop a multiple linear regression model~\cite{draper1998applied}. The dataset, consisting of 100 skull meshes, is divided into a training set (50 meshes) and a testing set (50 meshes) randomly. The model predicts the orbit curvature based on the skull's length $x$, width $y$, and height $z$ (all in millimeters (mm)). The resulting best-fit equation is: 
\begin{equation}
\text{Curvature}= 0.5653-0.0013 x - 0.0004 y - 0.0153 z. 
    \label{model}
\end{equation}

The regression coefficients reveal a striking anisotropy in how cranial dimensions constrain the orbit. The coefficient for skull height $z$ ($-0.0153$) is significantly greater than the others. This disparity indicates a functional decoupling of cranial modules. The high sensitivity to $z$ supports the architectural constraint \cite{marugan2006avian,marugan2022beyond}, which posits that the avian skull is a vertically integrated system where the neurocranium, orbit, and palate must be stacked within a confined architectural space. Consequently, the orbit appears to be vertically constrained between the neurocranium and palate; our model suggests that a reduction in height is strongly associated with a rounding of the orbital curvature. This comparison of regression coefficients is statistically robust, as the explanatory variables (length, width, and height) share identical units of measurement. We remark that as an empirical linear approximation, this model is only validated for the specific range of values encountered in our 100-specimen dataset: length $x$ ranges from 20.22 to 37.78 mm, width $y$ from 12.23 to 24.27 mm, and height $z$ from 11.03 to 23.01 mm. Extrapolating beyond these biological ranges, such as towards lengths approaching zero or excessively large dimensions, may yield non-physical curvature values. Therefore, this predictive model should be applied strictly within this bounded morphospace.

\begin{figure*}[t!]
    \centering
     \includegraphics[width=1\linewidth]{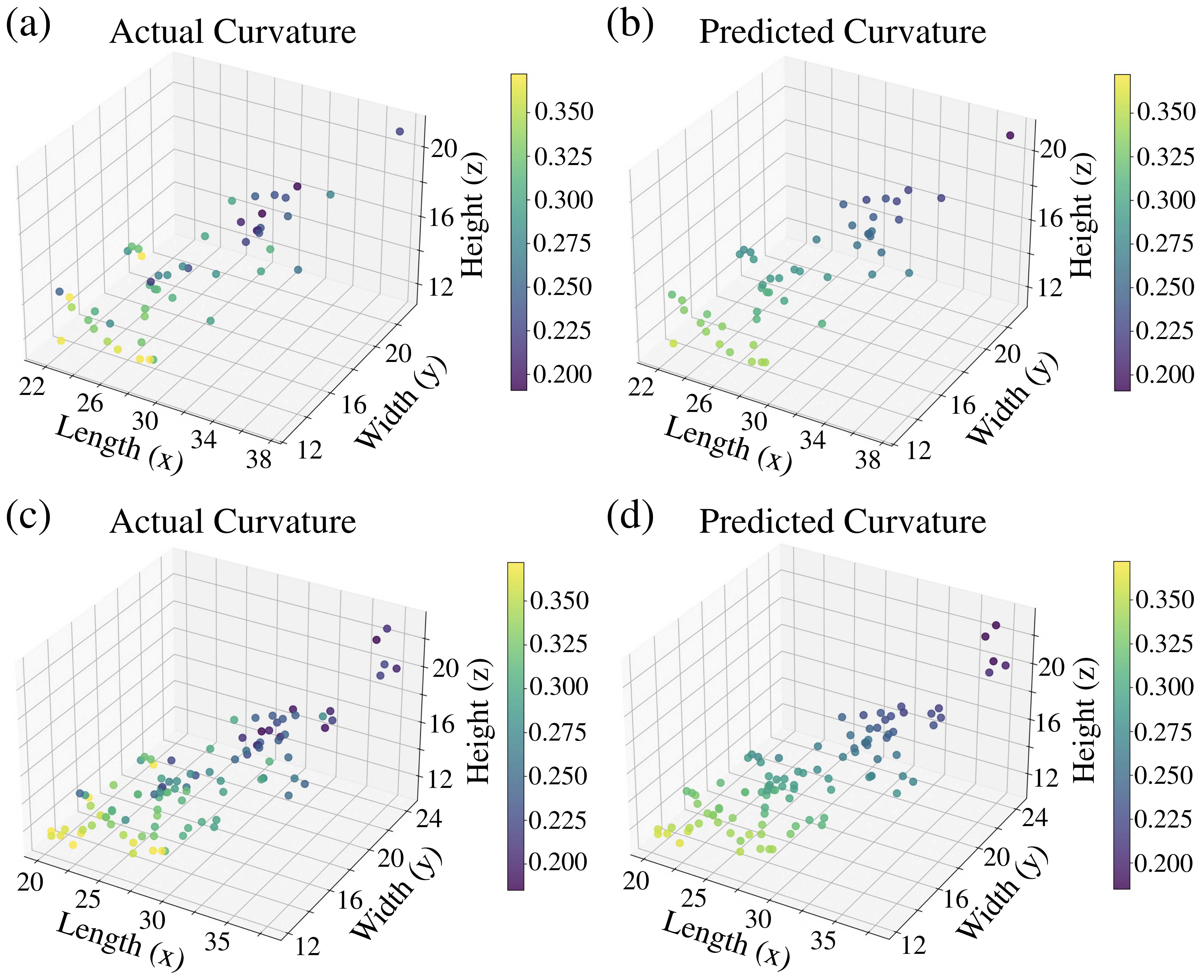}
    \caption{\textbf{Orbit curvature estimation from skull dimensions via our proposed model.} (a)~The 50 training samples with the coordinates of each point representing the skull dimensions ($x$, $y$, $z$) and the colors representing the actual (measured) orbit curvature. (b)~The data points color-coded with the predicted curvature obtained using the best-fit multiple linear regression model in Eq.~\eqref{model}. It can be observed that the curvature distributions in the two plots are highly consistent. (c)~The actual (measured) orbit curvature for all 100 specimens. (d)~The predicted curvature for all 100 specimens obtained using the model. }
    \label{fig:Actual_VS_Model_Curvature}
\end{figure*}

In Fig.~\ref{fig:Actual_VS_Model_Curvature}, we show the data points with the actual (measured) orbit curvature and the predicted curvature values obtained by the above best-fit model. It can be observed that the curvature distributions are highly consistent, providing strong evidence for the relationships between the curvature and skull dimensions. 

For a more quantitative analysis, we further evaluate the performance of the model by comparing the predicted curvature values with the actual curvature values from the testing set. The relative error is calculated using the following formula: 
\begin{equation}
    \text{Relative Error} = \frac{|\text{Actual curvature}-\text{Predict curvature}|}{\text{Actual curvature}} .
\end{equation}

The multiple linear regression model we developed demonstrates a high degree of effectiveness. Specifically, the model achieves an adjusted R-squared value of 0.8548 for the testing set, indicating that $85.48\%$ of the variance in orbit curvature can be explained by the skull's primary dimensions. A potential anatomical basis for this strong correlation is the highly integrated nature of the modern avian skull. As detailed by Plateau and Foth~\cite{plateau2020birds}, the evolutionary trend of bone fusion and integration created a structurally unified avian skull. This unity imposes strict rules on how the skull can grow and change shape, resulting in the strong, predictable allometric correlations between different features that we were able to model successfully. Also, the F-statistic probability is extremely low ($6.86\times 10^{-8}$), confirming that the model's predictive power is not due to random chance~\cite{james2013introduction}. The model's robustness is further validated by comparing its predictions to the actual curvature values of the 50 test samples. The average relative error for all 50 test samples is a low $6.35\%$, with a minimum error of $0.064\%$ and a maximum of $27.93\%$. A significant majority of the predictions (39/50, i.e., $78\%$) have a relative error below $10\%$, over half of the predictions (26/50, i.e., $52\%$) have a relative error below $5\%$, and a substantial portion (11/50, i.e., $22\%$) further exhibit a relative error below $1\%$. This demonstrates the model's robustness and its potential for use in sensitive scientific analyses. 

From a biological perspective, Eq.~\eqref{model} shows that generally a higher dimension size of the head results in a lower curvature. Eventually, by taking the inverse, we will have a larger orbit. This result parallels the usual developmental process of the skull in one of Darwin's finches by illustrating that the orbit's development is intrinsically linked to the growth of the entire skull, meaning a larger overall skull results from a growth process that also produces larger orbits~\cite{genbrugge2011ontogeny}. Besides, from a computational perspective, the regression model demonstrates the possibility of predicting complex geometric features from more easily obtainable linear skull measurements. This not only largely reduces the computational costs for shape quantification but also provides a promising direction for reconstructing or estimating complex features from damaged or incomplete specimens for further computational analyses.

\subsection{Correlation between Neurocranial Shapes and Other Features}

Also, using the proposed ellipsoid fitting algorithm for quantifying the neurocranium, we analyze the relationship between the ellipsoidal semi-axis lengths and other geometric features of the skull. 

In particular, a strong negative correlation is consistently observed between the lengths of the ellipsoid semi-axes ($a$, $b$, and $c$) and the orbit curvature. This indicates that as the overall size of the neurocranium increases (as modeled by the ellipsoid dimensions), the curvature of the orbit becomes quantitatively less curved. This inverse relationship is robust, evident in both Pearson and Spearman analyses (see SI Section~S3 for the detailed statistical results). Therefore, the algorithm not only quantifies size and shape independently but also captures the intrinsic geometric link between them, providing a quantitative framework for understanding how the scale of a cranial structure relates to its local topography. Moreover, in analyzing skull morphology, the semi-axes are more biologically informative than standard linear measurements because they directly represent the dimensions of the neurocranium. We find that a larger principal semi-axis, corresponding to a larger neurocranium, results in a smaller orbit curvature. This suggests that the spatial requirements of the neurocranium and orbit are tightly integrated during development, where cranial expansion accommodates orbital curvature changes through allometric scaling, consistent with the cranial integration patterns described by~\cite{tokita2017cranial}.

\subsection{Capturing Shape Information Independent of Size}

Besides enabling the estimation of orbit curvature and the neurocranial shapes from simple geometric measurements, it is natural to ask whether our framework can capture shape information independent of overall skull size for further shape analysis. Here, we define a dimensionless normalized curvature
\begin{equation}
\tilde{\kappa} \;=\; \frac{L}{\hat{r}},
\label{eq:norm_curv}
\end{equation}
where $L = (xyz)^{1/3}$ is the geometric mean of the bounding-box dimensions. It is easy to see that the normalized curvature $\tilde{\kappa}$ is invariant under uniform scaling. Moreover, we observe that the normalized curvature is effectively decoupled from size (Spearman $\rho = -0.07$, $p = 0.47$) yet still varies significantly across the genera in our dataset (one-way ANOVA, $F_{9,90} = 4.45$, $p < 0.001$, $\eta^2 = 0.285$), indicating that our framework captures genuine inter-genus shape variation isolated from size. In Fig.~\ref{fig:FIG_normalized_by_genus}, we further plot the normalized curvature $\tilde{\kappa}$ against the characteristic skull size $L$ for all 100 specimens. It can be observed that specimens from the same genus generally share close values of $\tilde{\kappa}$ regardless of the skull size $L$, indicating that the normalized curvature is useful for analyzing inter-genus shape variation.

\begin{figure}[t]
\centering
\includegraphics[width=1\textwidth]{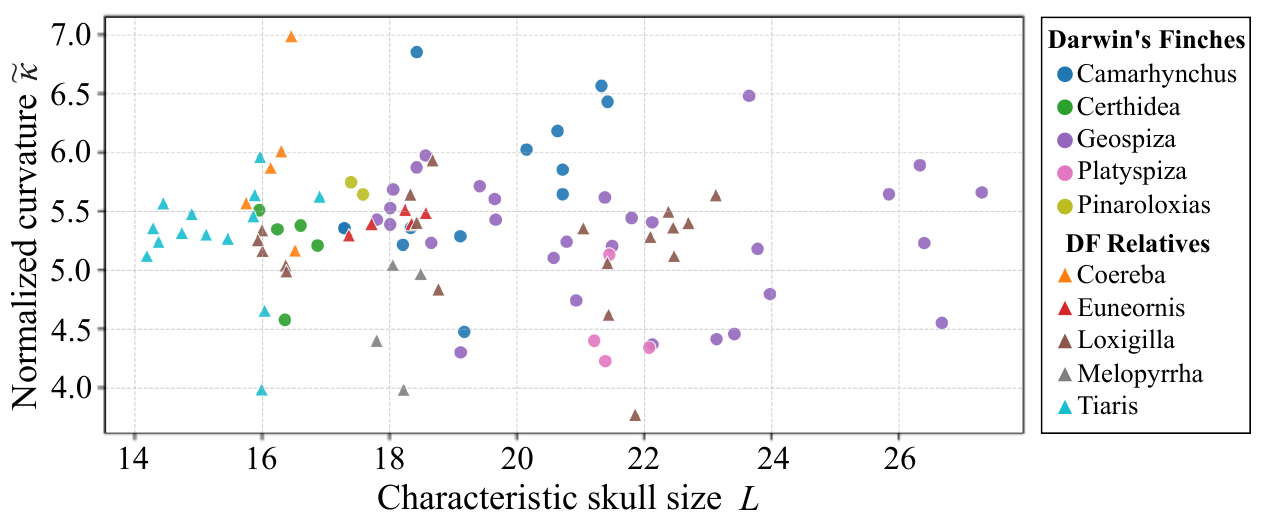}
\caption{\textbf{Scatter plot of the normalized curvature $\tilde{\kappa} = L/\hat{r}$ against the characteristic skull size $L = (xyz)^{1/3}$ for all 100 specimens}. Each marker represents one specimen (circles: Darwin's finch genera; triangles: DF relative genera) and is color-coded by genus. The normalized curvature is decoupled from size (Spearman $\rho = -0.07$, $p = 0.47$), confirming that the normalization isolates shape from size.} 
\label{fig:FIG_normalized_by_genus}
\end{figure}

To further examine this inter-genus variation using both components of our orbit and neurocranium estimation framework, we combine the orbit-derived normalized curvature $\tilde{\kappa}$ with another size-independent, neurocranium-derived descriptor obtained from the ellipsoid fitting, namely the aspect ratio $c/b$ of the two largest ellipsoidal semi-axes. As shown in Fig.~\ref{fig:morphospace_2panel}, each genus occupies a characteristic region of the resulting size-independent morphospace, whether $c/b$ is paired with the orbit curvature $\tilde{\kappa} = L/\hat{r}$ (panel~(a)) or with the skull aspect ratio $x/y$ (panel~(b)). Also, the neurocranium ratio $c/b$ is the single strongest genus discriminator among the size-independent descriptors examined (one-way ANOVA, $F_{9,90} = 10.40$, $p < 10^{-10}$, $\eta^2 = 0.510$). These results demonstrate that the parametric descriptors extracted by both components of our framework, orbit sphere fitting and neurocranium ellipsoid fitting, carry complementary, size-independent information that distinguishes genera.

\begin{figure}[t]
\centering
\includegraphics[width=\textwidth]{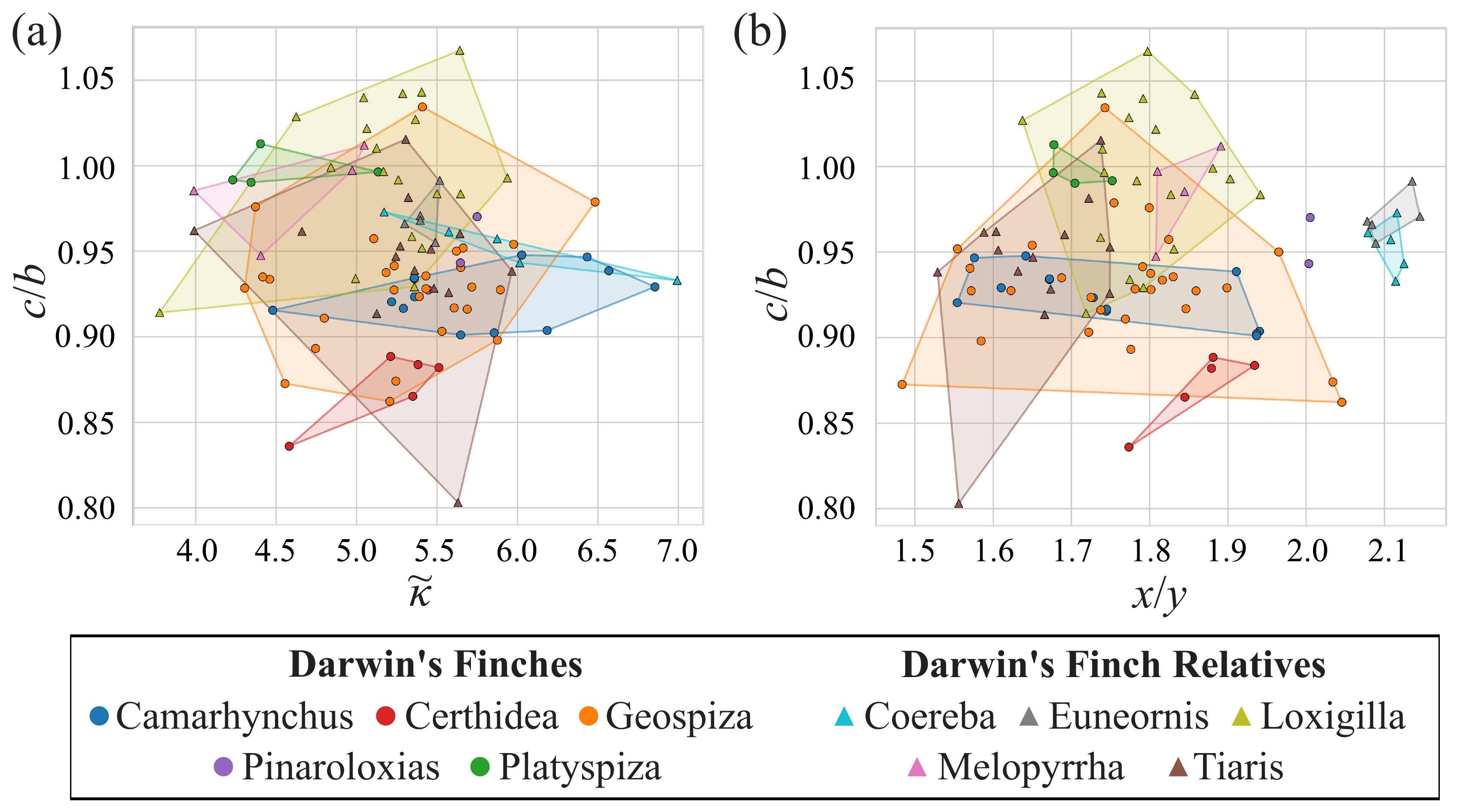}
\caption{\textbf{Size-independent inter-genus morphospaces combining both components of the framework.} Distribution of all 100 specimens of Darwin's finches (circles, blue palette) and their outgroup relatives (triangles, orange palette) in two size-independent morphospaces. Each point is one specimen. Both panels share the vertical axis: the neurocranium aspect ratio $c/b$ obtained from the ellipsoid fitting (one-way ANOVA, $F_{9,90} = 10.40$, $p < 10^{-10}$, $\eta^2 = 0.510$). (a)~Orbit-derived normalized curvature $\tilde{\kappa} = L/\hat{r}$ (one-way ANOVA, $F_{9,90} = 4.45$, $p < 0.001$, $\eta^2 = 0.285$) versus $c/b$, combining the orbit and neurocranium components of our pipeline. (b)~Skull aspect ratio $x/y$ (one-way ANOVA, $F_{9,90} = 15.65$, $p < 10^{-14}$, $\eta^2 = 0.610$) versus $c/b$. All three descriptors are dimensionless and statistically decoupled from overall skull size ($|\rho| \le 0.10$). Convex hulls indicate the region occupied by each genus (only applicable to genera with $>2$ specimens).} 
\label{fig:morphospace_2panel}
\end{figure}

\section{Discussion}

Our findings demonstrate that the integration of computational geometry and statistical modelling provides a robust framework for quantifying and estimating avian cranial morphology. By moving beyond traditional landmark-based approaches, this methodology successfully captures the relationships between primary skull dimensions, orbit curvature, and neurocranial geometry. In particular, our results has quantified the allometric relationship that skull height serves as a primary architectural constraint for orbital shape, suggesting that the avian cranium evolved as a tightly integrated structural unit. Altogether, our work not only offers insights into the evolutionary pressures shaping Darwin's finches but also establishes a scalable pipeline for the high-throughput phenotyping of large-scale museum collections, even when specimens exhibit significant damage.

\subsection{Biological Implications: Evidence for Anisotropic Cranial Integration}
Our regression analysis reveals that orbital curvature is not merely a function of isometric scaling but is shaped by distinct architectural constraints. The regression results highlight a particular sensitivity to vertical dimensions, where the coefficient for skull height ($-0.0153$) suggests a strong structural relationship between cranial altitude and orbital shape. The high sensitivity to skull height supports the architectural constraint hypothesis. The avian orbit is vertically compressed between the neurocranium and the palatal complex. Consequently, a reduction in skull height corresponds to a geometric rounding of the orbital curvature, suggesting a tight spatial integration where the orbit shape is maintained within the available vertical space defined by the neurocranium and palate.

\subsection{Methodological Robustness to Topological Noise}
A significant challenge in museum phenomics is the prevalence of taphonomic distortion and natural fenestration, which frequently renders specimens topologically complex. While advanced diffeomorphic mapping and parameterization techniques offer high-fidelity shape analysis, many of them are only applicable to shape data with a fixed topology. Applying high-dimensional geometric morphometrics and diffeomorphic mappings to raw museum scans often necessitates labor-intensive manual hole-filling or landmark placement, limiting scalability. In contrast, our primitive-fitting approach is designed as a specialized tool that parallels the morphometric revolution by prioritizing automation and high-throughput capability over fine-grained holistic shape analysis. By acting as a geometric low-pass filter, our iterative outlier-rejection algorithm isolates the functional envelope of the cranial features while ignoring topological defects. This allows for the automated phenotyping of damaged specimens that would otherwise be discarded or require extensive manual repair in standard geometric morphometrics pipelines.

\subsection{Applications in Digital Reconstruction and High-Throughput Phenomics}
The high predictive power of our model ($R^2 \approx 85\%$) establishes a rigorous framework for the parametric estimation of incomplete specimens. Since orbital geometry is tightly integrated with the robust linear axes of the skull, missing or fragmented orbital rims can be mathematically inferred from bounding box measurements alone. This provides a quantitative tool for parametric estimation in paleontology, enabling the recovery of morphometric data from damaged holotypes. Furthermore, by removing the requirement for manual landmarking, our fully unsupervised pipeline addresses the scalability bottleneck in comparative anatomy. As digitization initiatives such as MorphoSource~\cite{morphosource} and oVert (OpenVertebrate)~\cite{oVert} generate tens of thousands of 3D scans, our approach enables the data-driven exploration of macro-evolutionary trends across entire phylogenies, shifting the paradigm from hypothesis-driven analysis of small cohorts to high-throughput automated phenotyping.

\subsection{Broader Anatomical and Taxonomic Applicability}

While our automated pipeline offers significant advantages in processing speed and resilience to topological noise, it is inherently constrained by its geometric vocabulary. Unlike traditional geometric morphometrics, which can quantify arbitrary shapes using homologous landmarks, our method focuses only on anatomical features that can be meaningfully approximated by primitive geometries, such as spheres or ellipsoids.

However, within this geometric constraint, the method holds broad taxonomic utility. To verify its applicability beyond Darwin’s finches, we tested the pipeline on scans of other avian and mammalian skulls (see SI Section S4). Despite the extreme morphological divergence of the avian cranium and the entirely distinct architectural layout of the mammalian skulls, the automated algorithm successfully isolated and parameterized the target geometries. This indicates that the baseline assumptions of the model hold across distinct evolutionary radiations and vertebrate classes.

Beyond the specific avian orbital and neurocranial architecture presented in this study, this framework is readily adaptable to a wide array of questions in functional morphology. Many critical biological structures are governed by spherical or ellipsoidal physical constraints. Therefore, while our method does not replace the holistic shape capture of traditional geometric morphometrics, it provides a highly scalable, targeted tool for extracting specific, functionally relevant parametric geometries across the vertebrate tree of life.

\subsection{Limitations and Future Directions}

Note that the species analyzed in this study share a close evolutionary history. While our geometric analysis reveals strong correlations between cranial dimensions and orbital shape, these traits may also be influenced by phylogenetic inertia that has not been taken into consideration in our current approach. Future work incorporating phylogenetic comparative methods, such as Phylogenetic Generalized Least Squares~\cite{mundry2014statistical}, would be valuable to further disentangle the effects of shared ancestry from functional architectural constraints.

\bibliographystyle{ieeetr}
\bibliography{reference.bib}

\clearpage

%%%%%%%%%%%%%%%%%%%%%%%
\centerline{\LARGE\textbf{Supplementary Information}}
\appendix
\renewcommand\thefigure{S\arabic{figure}}    
\setcounter{figure}{0}
\renewcommand\thetable{S\arabic{table}}    
\setcounter{table}{0}
\renewcommand\thesection{S\arabic{section}}    
\setcounter{section}{0}

\section{Skull Dataset and Geometric Quantification}

In this work, we considered 100 specimens of Darwin's finches and their relatives from~\cite{tokita2017cranial,al2021geometry,mosleh2023beak} for our experiments. Table~\ref{tab:DF_list} and Table~\ref{tab:DFR_list} list the genus and species of each of the 100 specimens. 

As described in the main text, we first remeshed the 3D scans and enhanced the mesh quality for subsequent analysis. Note that while the relevant voxelization and smoothing procedures were also performed in~\cite{al2021geometry,mosleh2023beak},  these prior studies focused on the analysis of the beak shapes, and hence their remeshing setups only preserved the beak shape well while oversimplifying many other features of the skulls, such as the orbit (see Fig.~\ref{fig:SI_remeshing}, left). This makes the remeshing results from these prior studies not directly applicable to our analysis in this work. Therefore, we re-ran the mesh processing procedure on the raw 3D scans with a refined remeshing setup to improve the preservation of the skull features
(see Fig.~\ref{fig:SI_remeshing}, right). The new remeshed surfaces were then used for the subsequent geometric and statistical analysis.

Also, in the main text, we described our proposed sphere and ellipsoid fitting algorithms for extracting geometric measurements from skull meshes. In Fig.~\ref{fig:SI_sphere_examples}, we show additional examples of the sphere fitting results for orbit curvature quantification. In Fig.~\ref{fig:SI_ellipsoid_examples}, we show additional examples of the ellipsoid fitting results for the neurocranial geometry quantification. From these results, it can be observed that our proposed algorithms effectively capture the geometric features of a large variety of skull meshes.

\section{Statistical Analysis on Orbit Geometries}

As part of our statistical shape analysis, we studied the correlation between the skull dimensions (width $x$, length $y$, height $z$) and orbit curvature of Darwin's finches (DF), their relatives, and all of them. Below, we provide more detailed results of the statistical analysis.

In Table~\ref{tab:pearson_1overr}, we first calculate the Pearson correlation (Linear) between skull dimensions and the radius of the best-fit spheres to the orbit (i.e., 1/(orbit curvature)). Also, Table~\ref{tab:spearman_1overr} records Spearman correlation (Monotonic) between skull dimensions and 1/(orbit curvature). From both tables, we can see a strong positive correlation between the skull dimensions and the fitted sphere radius. 

In Table~\ref{tab:pearson_r} and Table~\ref{tab:spearman_r}, we further compute the Pearson correlation and Spearman correlation between the skull dimensions and the orbit curvature directly. As expected, we observe a strong negative correlation between skull dimensions and orbit curvature.

\section{Statistical Analysis on Neurocranial Geometries}

We also studied the correlation between the geometry of the neurocranium of the skull and the other geometric quantities of Darwin's finches (DF), their relatives, and all of them, using the ellipsoid fitting algorithm as described in the main text. 

In particular, we calculated the Pearson and Spearman correlations between the lengths of the ellipsoid semi-axes ($a$, $b$, $c$) and the orbit curvature of Darwin's finches, their relatives, and all of them. Table~\ref{tab:pearson_various_df} and Table~\ref{tab:spearman_various_df} record the results for Darwin's finches. Table~\ref{tab:pearson_various_dfr} and Table~\ref{tab:spearman_various_dfr} record the results for the relatives of Darwin's finches. Table~\ref{tab:pearson_various_both} and Table~\ref{tab:spearman_various_both} record the results for all of them. 

\section{Applicability to Other Biological Scans}

As mentioned in the main text, our method can be extended to a wider range of skulls. To illustrate this, we tested our pipeline on a parallel avian radiation, the Hawaiian honeycreepers (\emph{Vestiaria coccinea}, \emph{Telespiza cantans}, and \emph{Loxops coccineus}, \emph{Hemignathus wilsoni}, \emph{Melamprosops phaeosoma}, and \emph{Chlorodrepanis flava}) with 3D mesh data obtained from Tokita et al.~\cite{tokita2017cranial}. We also tested our pipeline on other mammalian skulls, specifically rodents of the genus \emph{Peromyscus} (\emph{Peromyscus spicilegus}, \emph{Peromyscus gossypinus}, and \emph{Peromyscus simulus}) with 3D mesh data obtained from the openVertebrate (oVert) project~\cite{oVert}. 

Here, it is noteworthy that Hawaiian honeycreepers exhibit highly diverse beak and skull shapes and thus serve as excellent test cases for our framework across different avian skulls. By contrast, the rodent skulls are notably different in terms of the overall skull geometry as well as the relative positions and sizes of the orbits, thereby serving as another set of excellent test cases for which the seed points may be harder to identify. As shown in Fig.~\ref{fig:SI_honeycreepers}, our algorithm successfully parameterized the orbits in all tested specimens without any clade-specific modifications, indicating that the geometric assumptions of our method generalize well beyond the Darwin's finch radiation analyzed in the main text.

\clearpage

\begin{figure}[h!]
    \centering
    \includegraphics[width=\linewidth]{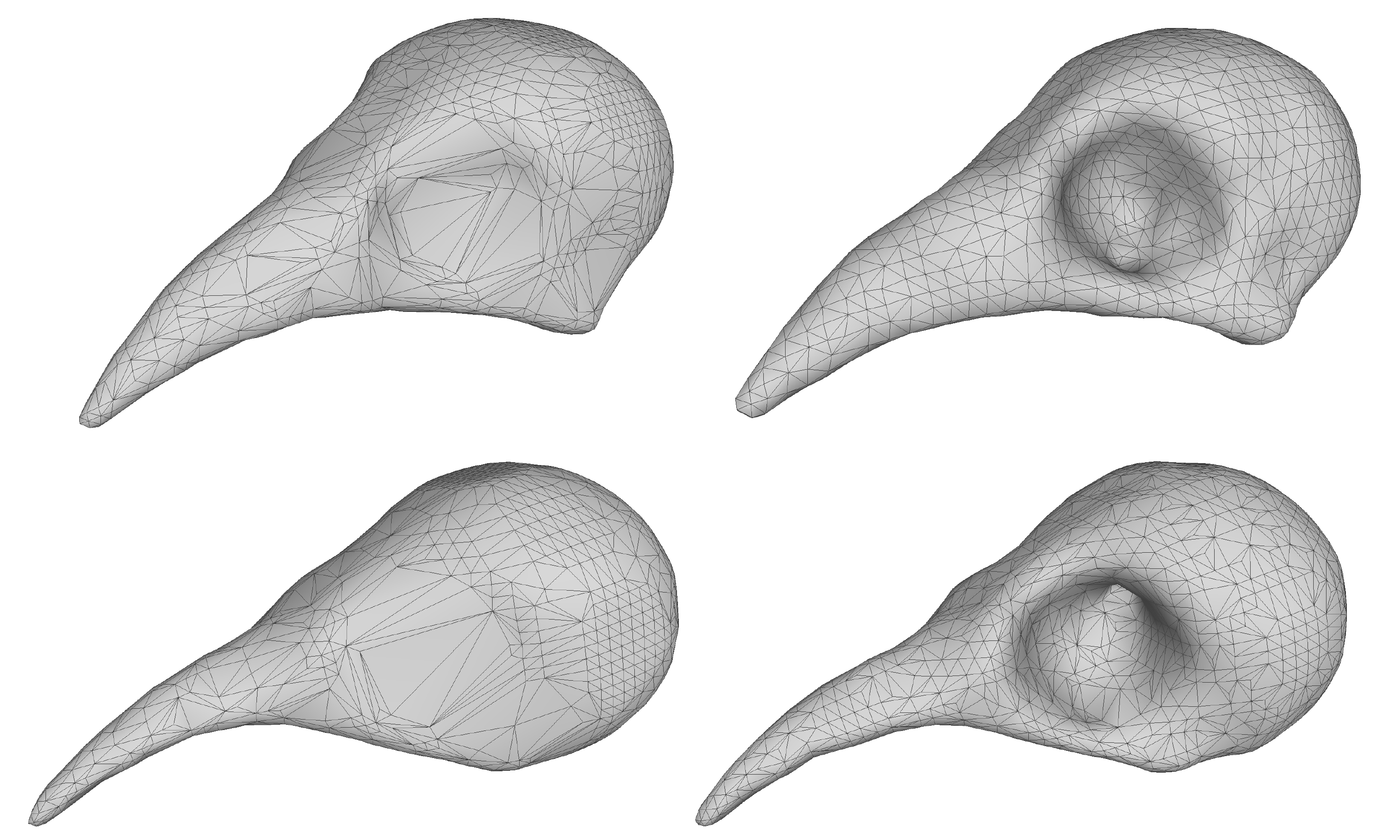}
    \caption{\textbf{Remeshing of the skull specimens in the dataset.} The left column shows the remeshing results by the prior studies~\cite{al2021geometry,mosleh2023beak}, the right column shows the new remeshing results done in this work. Top row: A specimen of Darwin's finches (\textit{Camarhynchus pallidus}). Bottom: A specimen of the relatives of Darwin's finches (\textit{Coereba flaveola}).}
    \label{fig:SI_remeshing}
\end{figure}

\begin{figure*}[h!]
    \centering
    \includegraphics[width=0.9\linewidth]{FIG_SI_sphere_examples.pdf}
    \caption{\textbf{More examples of the sphere fitting results for orbit curvature quantification achieved by our proposed algorithm.} The first three rows show examples of Darwin's finches. The last two rows show examples of the relatives of Darwin's finches. In each example, the seed point is highlighted in green, and the identified points for the fitting are highlighted in red. The optimal sphere is displayed in blue. The figures are not displayed to scale.}
    \label{fig:SI_sphere_examples}
\end{figure*}

\begin{figure*}[h!]
    \centering
    \includegraphics[width=0.9\linewidth]{FIG_SI_ellipsoid_examples.pdf}
    \caption{\textbf{More examples of the ellipsoid fitting results for neurocranial geometry quantification achieved by our proposed algorithm.} The first three rows show examples of Darwin's finches. The last two rows show examples of the relatives of Darwin's finches. In each example, the seed point is highlighted in green, and the identified points for the fitting are highlighted in red. The optimal ellipsoid is displayed in blue. The figures are not displayed to scale.}
    \label{fig:SI_ellipsoid_examples}
\end{figure*}

\begin{figure*}[h!]
    \centering
    \includegraphics[width=0.9\linewidth]{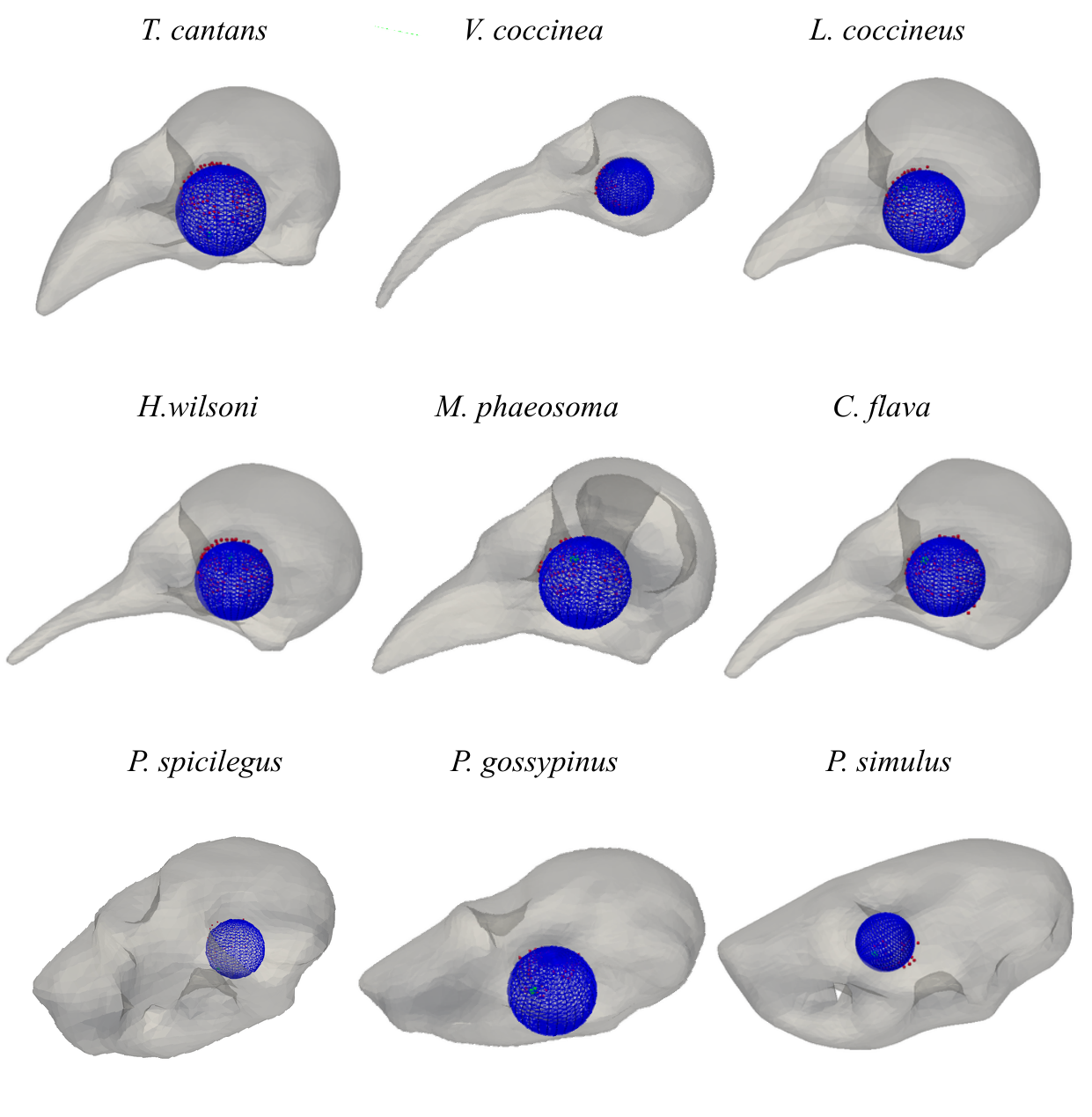}
    \caption{\textbf{Taxonomic extension of the automated primitive-fitting pipeline.} To test the broader applicability of the method beyond Darwin's finches, the algorithm was applied to highly divergent taxa. The algorithm successfully parameterized the orbital constraints across Hawaiian honeycreepers, including \emph{Telespiza cantans}, \emph{Vestiaria coccinea}, \emph{Loxops coccineus}, \emph{Hemignathus wilsoni}, \emph{Melamprosops phaeosoma}, and \emph{Chlorodrepanis flava}. The method was further tested on other mammalian skulls, including rodents in the genus \emph{Peromyscus} (\emph{Peromyscus spicilegus}, \emph{Peromyscus gossypinus}, \emph{Peromyscus simulus}), from which we observed that the method also successfully parameterized the geometric cavities. This demonstrates the method's robustness across entirely different vertebrate classes without the need for algorithmic re-parameterization.}
    \label{fig:SI_honeycreepers}
\end{figure*}

\begin{table}[t]
\centering
\resizebox{0.6\textwidth}{!}{
\begin{tabular}{|c|c|c|}
\hline
\textbf{Specimen ID} & \textbf{Genus} & \textbf{Species} \\
\hline
1 & \textit{Camarhynchus} (Tree Finch) & \textit{pallidus} \\
2 & \textit{Camarhynchus} (Tree Finch) & \textit{pallidus} \\
3 & \textit{Camarhynchus} (Tree Finch) & \textit{pallidus} \\
4 & \textit{Camarhynchus} (Tree Finch) & \textit{pallidus} \\
5 & \textit{Camarhynchus} (Tree Finch) & \textit{parvulus} \\
6 & \textit{Camarhynchus} (Tree Finch) & \textit{parvulus} \\
7 & \textit{Camarhynchus} (Tree Finch) & \textit{parvulus} \\
8 & \textit{Camarhynchus} (Tree Finch) & \textit{parvulus} \\
9 & \textit{Camarhynchus} (Tree Finch) & \textit{parvulus} \\
10 & \textit{Camarhynchus} (Tree Finch) & \textit{psittacula} \\
11 & \textit{Camarhynchus} (Tree Finch) & \textit{psittacula} \\
12 & \textit{Camarhynchus} (Tree Finch) & \textit{psittacula} \\
13 & \textit{Camarhynchus} (Tree Finch) & \textit{psittacula} \\
14 & \textit{Certhidea} (Warbler Finch) & \textit{fusca} \\
15 & \textit{Certhidea} (Warbler Finch) & \textit{fusca} \\
16 & \textit{Certhidea} (Warbler Finch) & \textit{olivacea} \\
17 & \textit{Certhidea} (Warbler Finch) & \textit{olivacea} \\
18 & \textit{Certhidea} (Warbler Finch) & \textit{olivacea} \\
19 & \textit{Geospiza} (Ground Finch) & \textit{conirostris} \\
20 & \textit{Geospiza} (Ground Finch) & \textit{conirostris} \\
21 & \textit{Geospiza} (Ground Finch) & \textit{conirostris} \\
22 & \textit{Geospiza} (Ground Finch) & \textit{conirostris} \\
23 & \textit{Geospiza} (Ground Finch) & \textit{conirostris} \\
24 & \textit{Geospiza} (Ground Finch) & \textit{conirostris} \\
25 & \textit{Geospiza} (Ground Finch) & \textit{conirostris} \\
26 & \textit{Geospiza} (Ground Finch) & \textit{difficilis} \\
27 & \textit{Geospiza} (Ground Finch) & \textit{difficilis} \\
28 & \textit{Geospiza} (Ground Finch) & \textit{fortis} \\
29 & \textit{Geospiza} (Ground Finch) & \textit{fortis} \\
30 & \textit{Geospiza} (Ground Finch) & \textit{fortis} \\
31 & \textit{Geospiza} (Ground Finch) & \textit{fortis} \\
32 & \textit{Geospiza} (Ground Finch) & \textit{fortis} \\
33 & \textit{Geospiza} (Ground Finch) & \textit{fuliginosa} \\
34 & \textit{Geospiza} (Ground Finch) & \textit{fuliginosa} \\
35 & \textit{Geospiza} (Ground Finch) & \textit{fuliginosa} \\
36 & \textit{Geospiza} (Ground Finch) & \textit{fuliginosa} \\
37 & \textit{Geospiza} (Ground Finch) & \textit{fuliginosa} \\
38 & \textit{Geospiza} (Ground Finch) & \textit{fuliginosa} \\
39 & \textit{Geospiza} (Ground Finch) & \textit{magnirostris} \\
40 & \textit{Geospiza} (Ground Finch) & \textit{magnirostris} \\
41 & \textit{Geospiza} (Ground Finch) & \textit{magnirostris} \\
42 & \textit{Geospiza} (Ground Finch) & \textit{magnirostris} \\
43 & \textit{Geospiza} (Ground Finch) & \textit{magnirostris} \\
44 & \textit{Geospiza} (Ground Finch) & \textit{scandens} \\
45 & \textit{Geospiza} (Ground Finch) & \textit{scandens} \\
46 & \textit{Geospiza} (Ground Finch) & \textit{septentrionalis} \\
47 & \textit{Geospiza} (Ground Finch) & \textit{septentrionalis} \\
48 & \textit{Platyspiza} (Vegetarian Finch) & \textit{crassirostris} \\
49 & \textit{Platyspiza} (Vegetarian Finch) & \textit{crassirostris} \\
50 & \textit{Platyspiza} (Vegetarian Finch) & \textit{crassirostris} \\
\hline
\end{tabular}
}
\caption{The list of specimens of Darwin’s finches considered in our study.}
\label{tab:DF_list}
\end{table}

\begin{table}[t]
\centering
\resizebox{0.6\textwidth}{!}{
\begin{tabular}{|c|c|c|}
\hline
\textbf{Specimen ID} & \textbf{Genus} & \textbf{Species} \\
\hline
51 & \textit{Platyspiza} (Vegetarian Finch) & \textit{crassirostris} \\
52 & \textit{Pinaroloxias} (Cocos Finch) & \textit{inornata} \\
53 & \textit{Pinaroloxias} (Cocos Finch) & \textit{inornata} \\
54 & \textit{Coereba} (Bananaquit) & \textit{flaveola} \\
55 & \textit{Coereba} (Bananaquit) & \textit{flaveola} \\
56 & \textit{Coereba} (Bananaquit) & \textit{flaveola} \\
57 & \textit{Coereba} (Bananaquit) & \textit{flaveola} \\
58 & \textit{Coereba} (Bananaquit) & \textit{flaveola} \\
59 & \textit{Euneornis} (Orangequit) & \textit{campestris} \\
60 & \textit{Euneornis} (Orangequit) & \textit{campestris} \\
61 & \textit{Euneornis} (Orangequit) & \textit{campestris} \\
62 & \textit{Euneornis} (Orangequit) & \textit{campestris} \\
63 & \textit{Euneornis} (Orangequit) & \textit{campestris} \\
64 & \textit{Loxigilla} (Bullfinch) & \textit{anoxanthus} \\
65 & \textit{Loxigilla} (Bullfinch) & \textit{anoxanthus} \\
66 & \textit{Loxigilla} (Bullfinch) & \textit{anoxanthus} \\
67 & \textit{Loxigilla} (Bullfinch) & \textit{anoxanthus} \\
68 & \textit{Loxigilla} (Bullfinch) & \textit{anoxanthus} \\
69 & \textit{Loxigilla} (Bullfinch) & \textit{noctis} \\
70 & \textit{Loxigilla} (Bullfinch) & \textit{noctis} \\
71 & \textit{Loxigilla} (Bullfinch) & \textit{noctis} \\
72 & \textit{Loxigilla} (Bullfinch) & \textit{noctis} \\
73 & \textit{Loxigilla} (Bullfinch) & \textit{portoricensis} \\
74 & \textit{Loxigilla} (Bullfinch) & \textit{portoricensis} \\
75 & \textit{Loxigilla} (Bullfinch) & \textit{portoricensis} \\
76 & \textit{Loxigilla} (Bullfinch) & \textit{portoricensis} \\
77 & \textit{Loxigilla} (Bullfinch) & \textit{portoricensis} \\
78 & \textit{Loxigilla} (Bullfinch) & \textit{violacea} \\
79 & \textit{Loxigilla} (Bullfinch) & \textit{violacea} \\
80 & \textit{Loxigilla} (Bullfinch) & \textit{violacea} \\
81 & \textit{Loxigilla} (Bullfinch) & \textit{violacea} \\
82 & \textit{Loxigilla} (Bullfinch) & \textit{violacea} \\
83 & \textit{Melopyrrha} (Cuban Bullfinch) & \textit{nigra} \\
84 & \textit{Melopyrrha} (Cuban Bullfinch) & \textit{nigra} \\
85 & \textit{Melopyrrha} (Cuban Bullfinch) & \textit{nigra} \\
86 & \textit{Melopyrrha} (Cuban Bullfinch) & \textit{nigra} \\
87 & \textit{Tiaris} (Grassquit) & \textit{bicolor} \\
88 & \textit{Tiaris} (Grassquit) & \textit{bicolor} \\
89 & \textit{Tiaris} (Grassquit)  & \textit{bicolor} \\
90 & \textit{Tiaris} (Grassquit)  & \textit{bicolor} \\
91 & \textit{Tiaris} (Grassquit)  & \textit{bicolor} \\
92 & \textit{Tiaris} (Grassquit) & \textit{canora} \\
93 & \textit{Tiaris} (Grassquit) & \textit{canora} \\
94 & \textit{Tiaris} (Grassquit) & \textit{canora} \\
95 & \textit{Tiaris} (Grassquit) & \textit{canora} \\
96 & \textit{Tiaris} (Grassquit) & \textit{canora} \\
97 & \textit{Tiaris} (Grassquit) & \textit{olivacea} \\
98 & \textit{Tiaris} (Grassquit) & \textit{olivacea} \\
99 & \textit{Tiaris} (Grassquit) & \textit{olivacea} \\
100 & \textit{Tiaris} (Grassquit) & \textit{olivacea} \\
\hline
\end{tabular}
}
\caption{The list of specimens of relatives of Darwin's finches considered in our study.}
\label{tab:DFR_list}
\end{table}

\clearpage

\begin{table*}[t]%
\centering
\begin{tabular*}{\textwidth}{@{\extracolsep\fill}lcccc@{\extracolsep\fill}}%
\toprule
 & \multicolumn{3}{c}{\textbf{1/(Orbit curvature)} of}\\
 & \textbf{Darwin's finches} & \textbf{DF relatives} & \textbf{All} \\
\midrule
\textbf{Length $x$}       & 0.711353 & 0.663881 &  0.722649 \\
\textbf{Width $y$}      & 0.773172 & 0.816721 & 0.805499 \\
\textbf{Height $z$}      & 0.760901 & 0.850506 & 0.819878 \\
\bottomrule
\end{tabular*}
\caption{Pearson correlation (Linear) between skull dimensions and 1/(orbit curvature).}
\label{tab:pearson_1overr}
\end{table*}

\begin{table*}[t]%
\centering
\begin{tabular*}{\textwidth}{@{\extracolsep\fill}lcccc@{\extracolsep\fill}}%
\toprule
 & \multicolumn{3}{c}{\textbf{1/(Orbit curvature)} of}\\
 & \textbf{Darwin's finches} & \textbf{DF relatives} & \textbf{All} \\
\midrule
\textbf{Length $x$}        & 0.739881 & 0.708161 & 0.758328 \\
\textbf{Width $y$}       & 0.801726 &  0.842623  & 0.832067 \\
\textbf{Height $z$}      & 0.774552 & 0.892692 & 0.844176 \\
\bottomrule
\end{tabular*}
\caption{Spearman correlation (Monotonic) between skull dimensions and 1/(orbit curvature).}
\label{tab:spearman_1overr}
\end{table*}

\begin{table*}[t]%
\centering
\begin{tabular*}{\textwidth}{@{\extracolsep\fill}lcccc@{\extracolsep\fill}}%
\toprule
 & \multicolumn{3}{c}{\textbf{Orbit curvature} of}\\
 & \textbf{Darwin's finches} & \textbf{DF relatives} & \textbf{All} \\
\midrule
\textbf{Length $x$}       & -0.7359881 & -0.681269 & -0.744344  \\
\textbf{Width $y$}     & -0.801726 & -0.812784 & -0.797486  \\
\textbf{Height $z$}      & -0.774552 & -0.864878 & -0.828820 \\
\bottomrule
\end{tabular*}
\caption{Pearson correlation (Linear) between skull dimensions and the orbit curvature.}
\label{tab:pearson_r}
\end{table*}

\begin{table*}[t]%
\centering
\begin{tabular*}{\textwidth}{@{\extracolsep\fill}lcccc@{\extracolsep\fill}}%
\toprule
 & \multicolumn{3}{c}{\textbf{Orbit curvature} of}\\
 & \textbf{Darwin's finches} & \textbf{DF relatives} & \textbf{All} \\
\midrule
\textbf{Length $x$}        & -0.642105 & -0.708161& -0.758328  \\
\textbf{Width $y$}    & -0.744534 & -0.842623 & -0.832067 \\
\textbf{Height $z$}      & -0.710526 & -0.892692 & -0.844176 \\
\bottomrule
\end{tabular*}
\caption{Spearman correlation (Monotonic) between skull dimensions and the orbit curvature.}
\label{tab:spearman_r}
\end{table*}

%==================================

\begin{table*}[t]
\centering
\begin{tabular*}{\textwidth}{@{\extracolsep\fill}lcccccc@{\extracolsep\fill}}%
\hline
\textbf{Variable} & \textbf{Length $x$} & \textbf{Width $y$} & \textbf{Height $z$} & \textbf{1/Curvature} & \textbf{Curvature} \\
\hline
\textbf{Ellipsoid semi-axis a} & 0.70 & 0.72 & 0.70 & 0.64 & -0.67 \\
\textbf{Ellipsoid semi-axis b} & 0.80 & 0.87 & 0.83 & 0.62 & -0.63 \\
\textbf{Ellipsoid semi-axis c} & 0.76 & 0.84 & 0.84 & 0.61 & -0.62 \\
\textbf{1/Curvature} & 0.71 & 0.77 & 0.76 & 1.00 & -0.98 \\
\textbf{Curvature} & -0.74 & -0.76 & -0.75 & -0.98 & 1.00 \\
\hline
\end{tabular*}%
\caption{Pearson correlation between various geometric quantities of Darwin's finches.}
\label{tab:pearson_various_df}
\end{table*}

\begin{table*}[t]
\centering
\begin{tabular*}{\textwidth}{@{\extracolsep\fill}lcccccc@{\extracolsep\fill}}%
\hline
\textbf{Variable} & \textbf{Length $x$} & \textbf{Width $y$} & \textbf{Height $z$} & \textbf{1/Curvature} & \textbf{Curvature} \\
\hline
\textbf{Ellipsoid semi-axis a} & 0.77 & 0.75 & 0.73 & 0.75 & -0.75 \\
\textbf{Ellipsoid semi-axis b} & 0.88 & 0.96 & 0.89 & 0.81 & -0.81 \\
\textbf{Ellipsoid semi-axis c} & 0.87 & 0.97 & 0.93 & 0.82 & -0.82 \\
\textbf{1/Curvature} & 0.74 & 0.80 & 0.77 & 1.00 & -1.00 \\
\textbf{Curvature} & -0.74 & -0.80 & -0.77 & -1.00 & 1.00 \\
\hline
\end{tabular*}

\caption{Spearman correlation between various geometric quantities of Darwin's finches.}
\label{tab:spearman_various_df}
\end{table*}

\begin{table*}[t]
\centering
\begin{tabular*}{\textwidth}{@{\extracolsep\fill}lcccccc@{\extracolsep\fill}}%
\hline
\textbf{Variable} & \textbf{Length $x$} & \textbf{Width $y$} & \textbf{Height $z$} & \textbf{1/Curvature} & \textbf{Curvature} \\
\hline
\textbf{Ellipsoid semi-axis a} & 0.52 & 0.69 & 0.68 & 0.66 & -0.62 \\
\textbf{Ellipsoid semi-axis b} & 0.76 & 0.95 & 0.94 & 0.85 & -0.84 \\
\textbf{Ellipsoid semi-axis c} & 0.81 & 0.98 & 0.98 & 0.86 & -0.87 \\
\textbf{1/Curvature} & 0.66 & 0.82 & 0.85 & 1.00 & -0.97 \\
\textbf{Curvature} & -0.68 & -0.81 & -0.86 & -0.97 & 1.00 \\
\hline
\end{tabular*}%

\caption{Pearson correlation between various geometric quantities of the relatives of Darwin's finches.}
\label{tab:pearson_various_dfr}
\end{table*}

\clearpage

\begin{table*}[t]
\centering
\begin{tabular*}{\textwidth}{@{\extracolsep\fill}lcccccc@{\extracolsep\fill}}%
\hline
\textbf{Variable} & \textbf{Length $x$} & \textbf{Width $y$} & \textbf{Height $z$} & \textbf{1/Curvature} & \textbf{Curvature} \\
\hline
\textbf{Ellipsoid semi-axis a} & 0.76 & 0.87 & 0.84 & 0.79 & -0.79 \\
\textbf{Ellipsoid semi-axis b} & 0.78 & 0.96 & 0.95 & 0.88 & -0.88 \\
\textbf{Ellipsoid semi-axis c} & 0.80 & 0.95 & 0.96 & 0.90 & -0.90 \\
\textbf{1/Curvature} & 0.71 & 0.84 & 0.89 & 1.00 & -1.00 \\
\textbf{Curvature} & -0.71 & -0.84 & -0.89 & -1.00 & 1.00 \\
\hline
\end{tabular*}%

\caption{Spearman correlation between various geometric quantities of the relatives of Darwin's finches.}
\label{tab:spearman_various_dfr}
\end{table*}

\begin{table*}[t]
\centering
\begin{tabular*}{\textwidth}{@{\extracolsep\fill}lcccccc@{\extracolsep\fill}}%
\hline
\textbf{Variable} & \textbf{Length $x$} & \textbf{Width $y$} & \textbf{Height $z$} & \textbf{1/Curvature} & \textbf{Curvature} \\
\hline
\textbf{Ellipsoid semi-axis a} & 0.62 & 0.68 & 0.69 & 0.65 & -0.66 \\
\textbf{Ellipsoid semi-axis b} & 0.81 & 0.92 & 0.89 & 0.74 & -0.75 \\
\textbf{Ellipsoid semi-axis c} & 0.60 & 0.89 & 0.90 & 0.73 & -0.75 \\
\textbf{1/Curvature} & 0.72 & 0.81 & 0.82 & 1.00 & -0.97 \\
\textbf{Curvature} & -0.74 & -0.80 & -0.83 & -0.97 & 1.00 \\
\hline
\end{tabular*}%
\caption{Pearson correlation between various geometric quantities of both Darwin's finches and their relatives.}
\label{tab:pearson_various_both}
\end{table*}

\begin{table*}[t]
\centering
\begin{tabular*}{\textwidth}{@{\extracolsep\fill}lcccccc@{\extracolsep\fill}}%
\hline
\textbf{Variable} & \textbf{Length $x$} & \textbf{Width $y$} & \textbf{Height $z$} & \textbf{1/Curvature} & \textbf{Curvature} \\
\hline
\textbf{Ellipsoid semi-axis a} & 0.77 & 0.85 & 0.83 & 0.77 & -0.77 \\
\textbf{Ellipsoid semi-axis b} & 0.84 & 0.97 & 0.95 & 0.84 & -0.84 \\
\textbf{Ellipsoid semi-axis c} & 0.86 & 0.95 & 0.96 & 0.86 & -0.86 \\
\textbf{1/Curvature} & 0.76 & 0.83 & 0.84 & 1.00 & -1.00 \\
\textbf{Curvature} & -0.76 & -0.83 & -0.84 & -1.00 & 1.00 \\
\hline
\end{tabular*}%

\caption{Spearman correlation between various geometric quantities of both Darwin's finches and their relatives.}
\label{tab:spearman_various_both}
\end{table*}

\end{document}